\documentclass[lettersize,journal]{IEEEtran}
\usepackage{amsmath,amsfonts}
\usepackage{algorithmic}
\usepackage{algorithm}
\usepackage{array}
\usepackage{textcomp}
\usepackage{stfloats}
\usepackage{url}
\usepackage{verbatim}
\usepackage{graphicx}
\usepackage{cite}
\usepackage{amsthm}
\usepackage{amsfonts}
\usepackage{amsmath} 
\usepackage{physics} 
\usepackage{algorithmic}  
\usepackage{algorithm} 
\usepackage{graphicx}
\usepackage{amssymb} 
\usepackage{color} 
\usepackage{bm}
\usepackage{epstopdf} 
\usepackage{subfigure}
\usepackage{booktabs}  
\usepackage[dvipsnames]{xcolor} 
\usepackage{cite} 
\usepackage{balance}
\usepackage{setspace}  
\usepackage{multirow} 
\usepackage{makecell} 
\usepackage{enumerate} 
\usepackage{subfigure} 
\usepackage{hyperref}

\newtheoremstyle{note}
{1pt}%
{1pt}%
{}%
{\parindent}%
{\rmfamily \bfseries}%
{:}%
{5pt}%
{}%
\theoremstyle{note} 

\newtheorem{ppn}{Proposition}
 
\newtheorem{remark}{Remark} 

\newtheorem*{pf}{Proof}  
\def\widebreve{\mathpalette\wide@breve}
\def\wide@breve#1#2{\sbox\z@{$#1#2$}%
	\mathop{\vbox{\m@th\ialign{##\crcr
				\kern0.08em\brevefill#1{0.8\wd\z@}\crcr\noalign{\nointerlineskip}%
				$\hss#1#2\hss$\crcr}}}\limits}
\def\brevefill#1#2{$\m@th\sbox\tw@{$#1($}%
	\hss\resizebox{#2}{\wd\tw@}{\rotatebox[origin=c]{90}{\upshape(}}\hss$}

\newcommand{\ppnref}[1]{\textbf{Proposition \ref{#1}}}
\newcommand{\algref}[1]{\textbf{Algorithm \ref{#1}}}

\makeatletter

\begin{document}
	
	\title{Soft Demodulator for Symbol-Level Precoding in Coded Multiuser MISO Systems}
	
	\author{Yafei Wang, \textit{Graduate Student Member}, \textit{IEEE}, Hongwei Hou, \textit{Graduate Student Member}, \textit{IEEE},\\ Wenjin Wang, \textit{Member}, \textit{IEEE}, Xinping Yi, \textit{Member}, \textit{IEEE}, Shi Jin, \textit{Senior Member}, \textit{IEEE}
	\thanks{Manuscript received xxx.}
		\thanks{Yafei Wang, Hongwei Hou, Wenjin Wang, Xinping Yi, and Shi Jin are with the National Mobile Communications Research Laboratory, Southeast University, Nanjing 210096, China (e-mail: wangyf@seu.edu.cn; 	
		hongweihou@seu.edu.cn; wangwj@seu.edu.cn; xyi@seu.edu.cn; jinshi@seu.edu.cn).}}
	
	\markboth{}%
	{Shell \MakeLowercase{\textit{et al.}}: A Sample Article Using IEEEtran.cls for IEEE Journals}
	
	
	\maketitle

	\begin{abstract}
		In this paper, we consider symbol-level precoding (SLP) in channel-coded multiuser multi-input single-output (MISO) systems. It is observed that the received SLP signals do not always follow Gaussian distribution, rendering the conventional soft demodulation with the Gaussian assumption unsuitable for the coded SLP systems. It, therefore, calls for novel soft demodulator designs for non-Gaussian distributed SLP signals with accurate log-likelihood ratio (LLR) calculation. To this end, we first investigate the non-Gaussian characteristics of both phase-shift keying (PSK) and quadrature amplitude modulation (QAM) received signals with existing SLP schemes and categorize the signals into two distinct types. The first type exhibits an approximate-Gaussian distribution with the outliers extending along the constructive interference region (CIR). In contrast, the second type follows some distribution that significantly deviates from the Gaussian distribution. To obtain accurate LLR, we propose the modified Gaussian soft demodulator and Gaussian mixture model (GMM) soft demodulators to deal with two types of signals respectively. Subsequently, to further reduce the computational complexity and pilot overhead, we put forward a novel neural soft demodulator, named pilot feature extraction network (PFEN), leveraging the transformer mechanism in deep learning. Simulation results show that the proposed soft demodulators dramatically improve the throughput of existing SLPs for both PSK and QAM transmission in coded systems.
	\end{abstract}
	

	%
	\IEEEpeerreviewmaketitle

	\section{Introduction}

	\IEEEPARstart{I}{n} multiuser multi-input multi-output (MU-MIMO) transmission, precoding is used to mitigate interference between users and increase spectral efficiency. Exploiting channel state information (CSI) at the transmitter/receiver, linear precoding schemes, such as the zero-forcing (ZF) precoding, have low computational complexity but cannot achieve the sum capacity in the finite signal-to-noise ratio (SNR) regime \cite{1468466, 1261332,1040325,1391204}.
	In contrast, symbol-level precoding (SLP), a nonlinear precoding method that leverages additional information from input data, goes beyond the performance achieved by linear precoding \cite{8359237, 9035662}. Unlike the conventional precoding schemes that aim to suppress interference and noise, SLP schemes introduce constructive interference into the received signals. 
	While conventional precoding schemes constrain the received signal to locate around the nominal constellations, SLP relaxes such constraint and allows the received signal to be distributed within a pre-designed constructive interference region (CIR), leading to improved performance \cite{8359237, 9035662, Li2021,4801492,5605266,6875291,7103338,7417066,8462190,7942010,8466792,8299553,8477154,8815429,8374931,9120670,shao2020minimum}. CIR is typically defined as the region more favorable to maximum-likelihood (ML) decision \cite{8299553}, based on which the two conventional optimizations, power minimization (PM) and signal-to-noise-plus-interference ratio (SINR) balancing, are redesigned for SLP transmission. The PM problem targets to minimize the total transmit power subject to SINR constraints
	\cite{6875291,7103338,7417066,8462190,7942010,8299553,8477154,7042789}, and SINR balancing aims to maximize the minimum received SINR with limited transmit power \cite{7103338,8466792,Li2021,8477154,7472286,8445846,7042789}.
	Specifically, the problems with the CIR for phase-shift keying (PSK) and quadrature amplitude modulation (QAM) symbols are respectively investigated in \cite{7103338} and \cite{7417066}, which are further formulated in various ways \cite{8466792,Li2021,8477154,8815429,8462190, 7942010}. 
	Additionally, the concept of CIR has been expanded to generic constellations, enabling a more generalized definition beyond the conventional PSK and QAM symbols \cite{8299553}.  
	Apart from PM and SINR balancing, symbol error rate (SER) minimization is also a crucial problem in SLP transmission \cite{8374931, shao2020minimum,9374958,9054488}, which generally further relaxes the constraint on the received signal distribution for lower SER. The SER minimization problem for PSK is considered in \cite{8374931}, while a deep-learning transceiver and an intelligent
	reflecting surface are introduced to minimize the SER for QAM in \cite{9054488} and \cite{shao2020minimum}, respectively.
	
	While attaining remarkable performance in uncoded systems, the freely distributed SLP signals impose significant challenges on soft demodulation for coded systems.
	Specifically, the received SLP signals are allowed to be freely distributed on the constellation map, exhibiting distinct non-Gaussian distributions for different SLP schemes \cite{7103338, 8374931,9054488}, and therefore different log-likelihood ratio (LLR) calculations for soft demodulation.
	Since conventional soft demodulators calculate LLR based on the assumption of Gaussian distribution, the presence of non-Gaussian signals poses challenges in computing the LLR for the coded system \cite{1495850,9435088,9035662}.
	However, to the best of our knowledge, little effort has been devoted to the investigation of soft demodulation for SLP with non-Gaussian received signals \cite{9035662}. Although some studies have explored the LLR calculation of non-Gaussian received signals \cite{9645175,9129383,9882269}, these methods are limited to scenarios with specific noise distribution, which is inapplicable to the SLP signals in coded systems. The main reason is that the distribution of SLP signals is quite different from previous studies, and more noticeably varies across different channel settings, constellation mappings, and SLP schemes.
	The above analysis raises a critical question: \textit{How to design effective and efficient soft demodulators for coded SLP transmission?}
	
	In this paper, we analyze the properties of the non-Gaussian received signal distribution and investigate new soft demodulators to obtain accurate LLR from the received signals. The major contributions of our work are summarized as follows:	
	\begin{itemize}
		\item 
		We first investigate the non-Gaussian characteristics of both PSK and QAM signals with existing SLP schemes and analyze their effect on LLR calculation. Based on the difference in received signal distributions, we categorize the non-Gaussian signals into two distinct types. The first type exhibits an approximate Gaussian distribution with the exception of a few probabilities extending along the CIR, and the second type has the distribution far deviated from the Gaussian distribution.
		
		\item We propose new soft demodulators for coded SLP systems with non-Gaussian received signals. Specifically, we propose a modified Gaussian soft demodulator for the first type of received signals. For the second type, we put forward a Gaussian mixture model (GMM) soft demodulator that approximates the non-Gaussian signal distribution by GMMs, with GMM parameters estimated from the received pilot signals. By exploiting the symmetry of the signal distributions corresponding to different constellation points, we propose a transform function, which reduce the number of unknown parameters and further decrease the pilot overhead. Simulation results demonstrate that GMM soft demodulators significantly improve the throughput of existing SLPs with PSK and QAM in coded systems.
		
		\item To avoid the iterative computation in GMM demodulator and the sharp performance decrease when pilot length is reduced, we propose the pilot feature extraction network (PFEN) soft demodulator, a deep-learning network trained to compute the optimal GMM parameters for received data signals from 
		limited-length pilot signals. The PFEN employs transformer modules with permutation invariance to comprehensively extract the features from the inputs, i.e., the received pilot signals and rescaling factor. Compared with the GMM demodulator, the PFEN demodulator requires lower complexity and fewer pilot overheads while maintaining excellent performance. 
	\end{itemize}
	
	This paper is structured as follows: In Section \ref{S2}, we introduce the system model and CIR. Section \ref{S4} analyzes the non-Gaussian distribution of the received signals. Section \ref{All Soft Demodulators} investigates modified Gaussian, GMM, and PFEN soft demodulators. Section \ref{results} reports the simulation results, and the paper is concluded in Section \ref{conclusion}. 
	
	{\textit{Notation}}: $(\cdot)^{-1}, (\cdot)^T, (\cdot)^H$ denote the transpose and the transpose-conjugate operations, respectively. $x$, ${\bf x}$, and ${\bf X}$ respectively denote a scalar, column vector, and matrix. $\real(\cdot)$ and $\imaginary(\cdot)$ represent the real and imaginary part of a complex scalars, vector or matrix. $j=\sqrt{-1}$ denote imaginary unit. $\in$ denotes belonging to a set, and $\sim$ denote being distributed as. The expression $\mathcal{C}\mathcal{N}(\mu, \sigma^2)$ represents circularly symmetric Gaussian distribution with mean $\mu$ and variance $\sigma^2$. ${\mathbb{R}}^{M\times N}$ and ${\mathbb{C}}^{M\times N}$ respectively denote the sets of $M\times N$ real- and complex-valued matrices. $\mathcal{A}\backslash\mathcal{B}$ means objects that belong to set $\mathcal{A}$ and not to $\mathcal{B}$. $\nabla f$ denotes gradient of function $f(\cdot)$. $|\mathcal{A}|$ represents the cardinality of set $\mathcal{A}$.  ${\rm sign}(\cdot)$ denotes the sign function. $\angle(x)$ denotes the angle of complex scalar $x$. $\lfloor\cdot\rfloor$ represents the floor function. ${\bf I}_{K}$ denotes $K\times K$ identity matrix. $\left \|\cdot\right \|_{2}$ denotes $l_2$-norm. ${\rm det}({\bf A})$ represents the determinant of matrix ${\bf A}$.
	
	\section{System and Signal Model}\label{S2}
	
	\subsection{System Model}\label{system model}
	Consider an MU-MISO downlink system where an $N$-antenna base station (BS) transmits the signal to $K$ single-antenna user equipment (UE). We assume block flat fading channels where the channel coefficients remain constant for a
	coherence interval of $L$ symbol durations. The channel between BS and the $k$-th UE is denoted as  ${\bf h}_k\in{\mathbb{C}}^{N\times 1}$. The channel matrix ${\bf H} = \left[{\bf h}_1,{\bf h}_2 ..., {\bf h}_K\right]^T$ is assumed to be available at the BS.
	
	We consider the SLP system where the received signal of $k$-th UE at $l$-th symbol duration is
	\begin{equation}
		\label{E1}
		{y}_k[l] = {\bf h}^T_k{\bf x}[l] + n_k[l],\;\forall k \in {\mathcal{ K}},
	\end{equation}
	where ${\mathcal{ K}}=\left\{1, 2, ..., K\right\}$, $n_k[l]\sim \mathcal{C}\mathcal{N}(0, \sigma^2)$ denotes the additive noise at the $k$-th UE and $\sigma^2$ represents the noise variance. ${\bf x}[l]\in{\mathbb{C}}^{N\times 1}$ is the transmit signal vector encoded by the symbol-level precoder for the transmission of ${\bf s}[l]$, which contains $K$ independent QAM or PSK symbols
	\begin{align}
		{\bf s}[l] = \left[s_{1}[l],s_{2}[l] ..., s_{K}[l]\right]^T,
	\end{align}
	where $s_{k}[l]$ is the symbol desired by the $k$-th UE. These symbols are drawn from the constellation set
	${\mathcal{V}}=\{v_q\}_{q\in {\mathcal{ Q}}}$, where ${\mathcal{ Q}}=\{1, ..., Q\}$ and $v_q$ represents $q$-th type of constellation point.
	
	Due to the distinct SLP schemes to obtain ${\bf x}[l]$, the distribution of the received signal varies accordingly. We start with the mostly adopted SLP, i.e., CI-based SINR balancing (CISB) \cite{Li2021, 8466792, 8815429}, and then extend the results to other schemes. The transmit signal can be written as \cite{8815429,7042789}
	\begin{align}
		{\bf x}[l]={\gamma}[l]\!\cdot\!{\bf H}^{\dagger}{\tilde{\bf s}}[l],
		\label{tx signal CI}
	\end{align}
	where ${\gamma}[l]=\sqrt{\frac{P_{\rm T}[l]}{\|{\bf H}^{\dagger}{\tilde{\bf s}}[l]\|^2_2}}$, ${\bf H}^{\dagger}={\bf H}^H({\bf H}{\bf H}^H)^{-1}$, and $P_{\rm T}[l]$ represents the transmit power. ${\tilde{\bf s}}[l]=\left[{\tilde s}_{1}[l],{\tilde s}_{2}[l] ..., {\tilde s}_{K}[l]\right]^T$ denotes the target signal vector, where ${\tilde s}_{k}[l]$ is constrained in the CIR of $s_k[l]$ and optimized to maximize ${\gamma}[l]$ \cite{8466792}. Substituting \eqref{tx signal CI} into \eqref{E1}, we have
	\begin{align}
	y_{k}[l] = {\gamma}[l]{\tilde s}_{k}[l] + n_k[l].
	\label{y SLP}
	\end{align}
	
	\begin{figure}[t]
		\centering
		\includegraphics[width=2.7in]{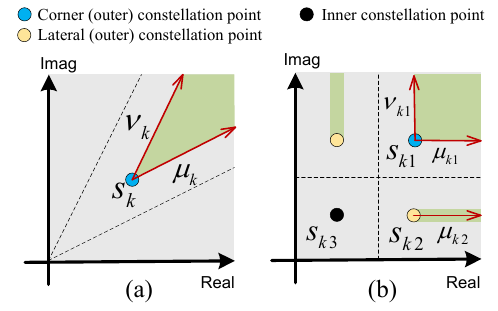}
		\caption{CIRs and their boundary vectors of (a) PSK and (b) QAM.}
		\label{CI_MMSE_CIR}
	\end{figure}
	
	\begin{figure*}[h]
		\centering
		\includegraphics[width=6in]{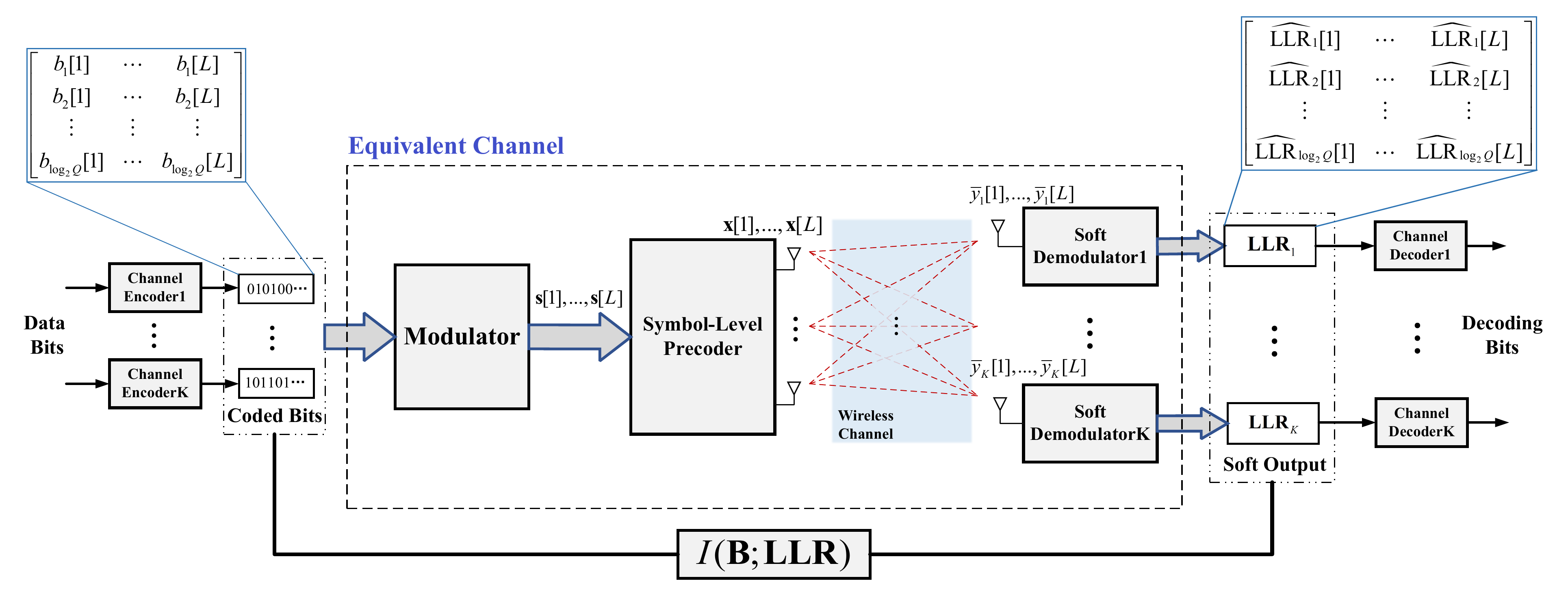}
		\caption{Coded wireless communication system with SLP.}
		\label{Coded System}
	\end{figure*}
	
	The CIR ${\mathcal D}_k[l]$ is the modulation-specific region where the interference component of the received signal is CI \cite{9035662,8299553,Li2021}. Here CI refers to the interference that pushes the noise-free received signal away from the ML decision boundaries.
	As illustrated in Fig. \ref{CI_MMSE_CIR}, the CIR (green areas) allows CI to extend outer real and imaginary parts of the signal while maintaining the
	performance of inner ones \cite{9035662}. It is worth noting that, for QAM, there are two types of constellation points, inner and outer (lateral and corner) ones, while the PSK only has outer (corner) ones.
	
	\subsection{Signal Model for Demodulation}
	When multi-level QAM is employed, the received signals are required to be scaled for correct demodulation \cite{li2020symbol}, and the signal to be demodulated turns to ${\bar{y}}_k[l] = {{y}_k[l]}/{\gamma[l]}={\tilde s}_k[l]+n_k[l]/{\gamma[l]}$. In practical block transmission, BS needs to broadcast $\gamma[l]$ to UEs at a symbol level, which results in excessive signaling overheads. 
	One method to facilitate practical demodulation is that BS employs the power allocation scheme in \cite{li2020symbol} to unify $\gamma[l]$ in a transmission block:
	\begin{align}
		&{\bar \gamma} = \sqrt{\frac{\sum_{l=1}^{L}P_{\rm T}[l]}{{\sum_{l=1}^{L}\frac{P_{\rm T}[l]}{\gamma^2[l]}}}},\ {\bar{\bf x}}[l]= \frac{{\bar \gamma}}{\gamma[l]}{\bf x}[l],\ \forall l \in {\mathcal{ L}},
		\label{PA_SLP2}
	\end{align}
	where ${\mathcal{ L}}=\{1,2,...,L\}$. Since PSK does not modulate the amplitude, rescaling and power allocation are optional. According to \eqref{y SLP} and \eqref{PA_SLP2}, the signal to be demodulated can be expressed as
	\begin{align}
	&{\rm QAM}: {\bar{y}}_k[l]\!=\!{\tilde s}_k[l]+\frac{n_k[l]}{{\bar \gamma}},\label{QAM demodulation}\\
	&{\rm PSK}: {\bar{y}}_k[l]\!=\!
	\begin{cases}
	{\tilde s}_k[l]+\frac{n_k[l]}{{\bar \gamma}},\ {\rm  WR}\\
	{{\gamma[l]}}{\tilde s}_k[l] + n_k[l],\ {\rm  WOR}
	\end{cases},
		\label{PSK demodulation}
	\end{align}
	where we use `WR' and `WOR' to represent `with rescaling' and `without rescaling' for convenience, respectively.
	
	When the ML decision rule is applied in an uncoded system, the symbol in ${\mathcal V}$ that is closest to the received signal in Euclidean distance is chosen for demodulation \cite{4282117}. By maximizing $\gamma[l]$, SLP increases the received SNR and achieves a lower symbol error rate (SER) than conventional precoding schemes \cite{7103338}. In addition, the CI in ${\tilde s}_k[l]$ pushes the signal away from the ML decision boundary and further reduces the SER. It is worth noting that these advantages brought by SLP are attainable for uncoded systems.
	When it comes to the coded systems with channel coding, however, it calls for new treatments for soft demodulation.
	
	\subsection{Soft Demodulation}
	Fig. \ref{Coded System} shows the multi-user coded systems with SLP, which includes channel encoders, channel (soft) decoders, and soft demodulators.
	Channel coding is an essential technique in practical communication systems, which usually employs an iterative soft decoder to approximate the optimal performance \cite{arora2020survey}. The LLR, as the output of the soft demodulator, is the sole input to the iterative soft decoder, whose exactness determines the performance of the iterative soft decoder \cite{1057683, 8630510}.
	
	For brevity of real representation, we define{\footnote{We focus on the demodulator of a single UE, so the index $k$ is temporarily omitted in ${\bf y}[l]$.}}
			\begin{align}
				{\bf{v}}_q = \begin{bmatrix}
					\real\left(v_q\right)\\
					\imaginary\left(v_q\right)
				\end{bmatrix}
				,\ {{\bf y}}[l] = 
				\begin{bmatrix}
					\real\left({\bar { y}}_k[l]\right)\\
					\imaginary\left({\bar { y}}_k[l]\right)
				\end{bmatrix}.
			\end{align}
	In general, symbols in ${\mathcal{V}}$ are transmitted with equal probability, and the LLR can be acquired from below \cite{485714}
			\begin{align}
				{\rm LLR}_i[l]= {\rm ln}\frac{\sum\limits_{{\bf{v}}_q\in{{\mathcal{S
							}}^{+}_i}}f_{{ Y}}({{\bf y}}[l]|{\bf{v}}_q)}{\sum\limits_{{\bf{v}}_q\in{{\mathcal{S
							}}^{-}_i}}f_{{ Y}}({{\bf y}}[l]|{\bf{v}}_q)},
				\label{E6b}
			\end{align}
	where ${\rm LLR}_i[l]$ denotes the LLR for $i$-th coded bit of ${{\bf y}}[l]$, and $f_{{ Y}}({{\bf y}}|{\bf{v}}_q)$ is the likelihood function \cite{8630510}. ${\mathcal{S}}^{+}_i$ and ${\mathcal{S}}^{-}_i$ are the sets of symbols whose $i$-th bit is 1 and 0, respectively. 
	
	\section{{The Effect of Non-Gaussian Received Signals on Demodulation}}\label{S4}	
	
	The conventional Gaussian soft demodulator works under the assumption that $f_{{ Y}}({{\bf y}}|{\bf{v}}_q)$ is the probability density function (PDF) of a complex Gaussian distribution and can be expressed as \cite{8283609}
	\begin{align}
		{f}_{{ Y}}({{\bf y}}|{\bf{v}}_q)=\frac{\exp\left({\!-\frac{\|{{\bf y}}-{\bf{v}}_q\|_2^2}{{\sigma}_{\rm s}^2}}\right)}{\pi{\sigma}_{\rm s}^2},\ \forall q \in {\mathcal{ Q}},
		\label{Gaus PDF}
	\end{align}
	where ${\sigma}^2_{\rm s}$ is the variance. However, the received signals of SLP schemes unlikely follow the Gaussian distribution, due to the presence of CI in the received signal. In particular, the received signals of SLP exhibit a departure from this assumption, i.e., ${f}_{{ Y}}({{\bf y}}|{\bf{v}}_q)$ with SLP is not Gaussian distributed. 
	
\begin{figure}[htbp]
		\centering
		\includegraphics[width=3in]{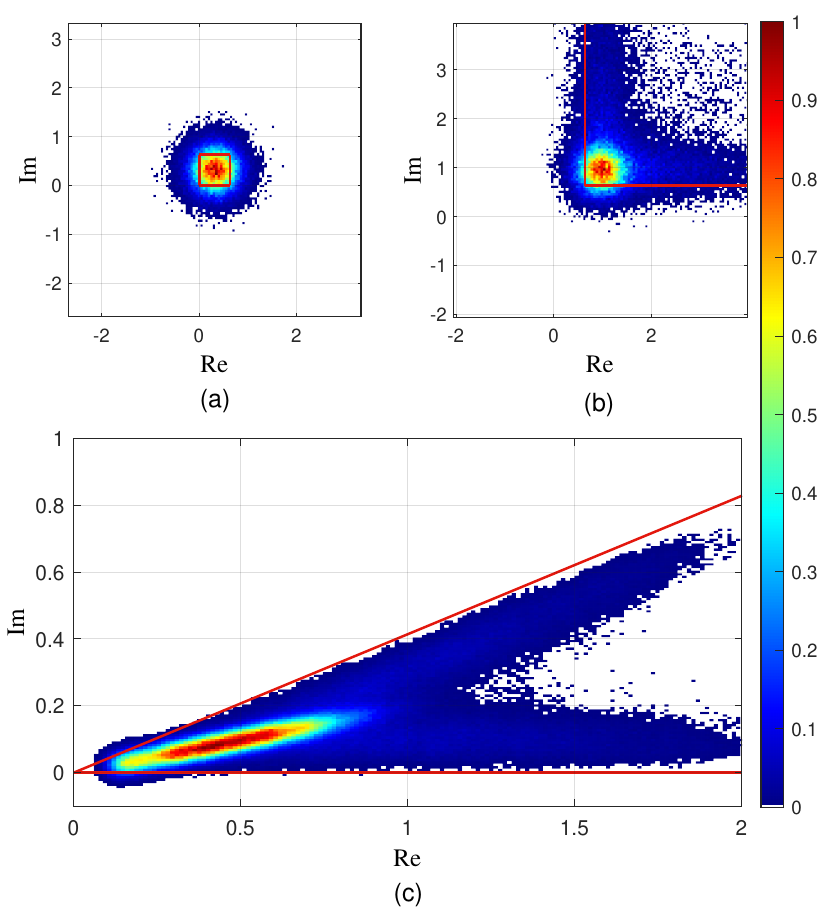}
		\caption{Normalized $f_{{ Y}}({{\bf y}}[l]|{\bf{v}}_q)$ with CISB where ${\bf{v}}_q$ is (a) inner constellation point from 16QAM. (b) corner constellation point from 16QAM. (c) constellation point from 16PSK (WOR). The red lines represent the decision boundaries.}
		\label{recieved distribution_small}
	\end{figure}

	In the scenario of $N=K=8$, Rayleigh channel, and ${\rm SNR}=P_{\rm T}/\sigma^2=20{\rm dB}$, Fig. \ref{recieved distribution_small} shows $f_{{ Y}}({{\bf y}}|{\bf{v}}_q)$ of three types of constellation points with CISB. We approximate $f_{{ Y}}({{\bf y}}|{\bf{v}}_q)$ using Monte Carlo simulations combined with statistical histograms, where the number of points within each square region is divided by their maximum value for normalization. In Fig. \ref{recieved distribution_small} (a), the distribution of $f_{{ Y}}({{\bf y}}|{\bf{v}}_q)$ exhibits a uniform circular pattern. In Fig. \ref{recieved distribution_small} (b), apart from the central circular pattern, $f_{{ Y}}({{\bf y}}|{\bf{v}}_q)$ further extends along both sides of the decision boundary. In Fig. \ref{recieved distribution_small} (c), the central distribution appears in a linear pattern, accompanied by extended portions. Evidently, $f_{{ Y}}({{\bf y}}|{\bf{v}}_q)$ does not adhere strictly to a Gaussian distribution.
	
	\subsection{Properties of Non-Gaussian Received Signals}
	\label{Signals Properties}
	
	In this subsection, we focus on the received signals of SLP and analyze the impact of their non-Gaussian distribution on the conventional soft demodulator.
	
	According to \eqref{E6b}, computing LLR requires the PDF $f_{{ Y}}({{\bf y}}|{\bf{v}}_q)$. As shown in \eqref{PSK demodulation}, while $n_k[l]\sim \mathcal{C}\mathcal{N}(0, \sigma^2)$ and ${\bar \gamma}$  can be seen as a constant when $L$ is long enough, $f_{{ Y}}({{\bf y}}|{\bf{v}}_q)$ is determined by the distribution of ${\tilde s}_k[l]$ and $\gamma[l]{\tilde s}_k[l]$. ${\tilde {\bf s}}[l]$ can be obtained by solving the following problem \cite{8815429,9910472}:
	\begin{align}
	\begin{split}
	\min\limits_{{\tilde {\bf s}}}\ \|{\bf H}^{\dagger}{\tilde{\bf s}}\|^2_2\ \  {\rm s.t.}\ {\tilde s}_k\in \mathcal{D}_k[l], \forall k\in{\mathcal{K}}.
	\end{split}
	\label{power minimization}
	\end{align}
	Since $\mathcal{D}_k[l]$ depends on $s_k[l]$, the above expression reveals that the change of $s_{k'}[l],k'\in{\mathcal{K}}\backslash k$ affects the optimal ${\tilde s}_k[l]$ of a fixed $s_k[l]$. We define
	\begin{align}
	&{\bf s}_{{\mathcal{K}}\backslash k}[l] \!\triangleq\! \begin{bmatrix}
				s_1[l]&...&s_{k-1}[l]&s_{k+1}[l]&...&s_K[l]
			\end{bmatrix}^T,\\
	&\mathcal{V}^{K} \!\triangleq\! \{{\bf s}:{\bf s}\!=\!\begin{bmatrix}
			s_{1} & s_{2} & \cdots & s_{K}
		\end{bmatrix}^T,\ \forall s_{k}\!\in\!\mathcal{V}\},
	\end{align}
	where ${\bf s}_{{\mathcal{K}}\backslash k}[l]$ represents the vector composed of the parts of ${\bf s}[l]$ except for $s_k[l]$. We have ${\bf s}_{{\mathcal{K}}\backslash k}[l]\in\mathcal{V}^{K-1}$, and user symbols in ${\bf s}_{{\mathcal{K}}\backslash k}[l]$ has $Q^{K-1}$ combinations, i.e., $|\mathcal{V}^{K-1}|=Q^{K-1}$. Based on the definition of ${\bf s}_{{\mathcal{K}}\backslash k}[l]$, we denote ${\tilde {\bf s}}^{\star}_{{\boldsymbol{\alpha}}|v_q}$ as the optimal solution of problem \eqref{power minimization} with $s_k[l]=v_q$ and ${\bf s}_{{\mathcal{K}}\backslash k}[l]={\boldsymbol{\alpha}}$, where it can be conlucded from the constraint in \eqref{power minimization} that $[{\tilde {\bf s}}^{\star}_{{\boldsymbol{\alpha}}|v_q}]_k\in \mathcal{D}_k[l]$. According to the definition of $\gamma[l]$, we further define $\gamma_{{\boldsymbol{\alpha}}|v_q}=\sqrt{{P_{\rm T}[l]}/{\|{\bf H}^{\dagger}{\tilde {\bf s}}^{\star}_{{\boldsymbol{\alpha}}|v_q}\|^2_2}}$. Assume constellation points in ${\mathcal V}$ are tranmitted with equal probability. Therefore, we have
	\begin{align}
			\begin{split}
			&P({\tilde s}_k[l]\!=\![{\tilde {\bf s}}^{\star}_{{\boldsymbol{\alpha}}|v_q}]_k |s_k[l]\!=\!v_q)=P\left(\gamma[l]\!=\!\gamma_{{\boldsymbol{\alpha}}|v_q}|s_k[l]\!=\!v_q\right)\\
			&=\frac{1}{Q^{K-1}},\ {\boldsymbol{\alpha}}\in{\mathcal V}^{K-1},
			\end{split}
			\label{discrete PDF}
	\end{align}
	based on which the distribution of ${\tilde s}_k[l]$ and $\gamma[l]{\tilde s}_k[l]$ can be obtained. 
	
	Considering that the solution of problem \eqref{power minimization} is a non-intuitive process, we analyze $f_{{ Y}}({{\bf y}}|{\bf{v}}_q)$ with the support of Fig. \ref{recieved distribution_small}:
	\begin{itemize}
	\item \textbf{Inner QAM constellation point:} According to Fig. \ref{CI_MMSE_CIR}, when $s_k[l]=v_q$ and $v_q$ is the inner constellation point, we have ${\tilde s}_k[l]=s_k[l]$, based on which ${\bar{y}}_k[l]$ with rescaling in \eqref{QAM demodulation}, \eqref{PSK demodulation} is expressed as 
	\begin{align}
	\overset{\qquad\qquad\qquad\qquad{\text{Gaussian random variable}}}{{\bar{y}}_k[l]\ \ =\ \  s_k[l]\ \ +\ \ \overbrace{{n_k[l]}/{{\bar \gamma}}}}.
	\label{inner distribution with rescaling}
	\end{align}
	${\bar{y}}_k[l]$ is a Gaussian random variable with mean $v_q$, and its distribution is shown in Fig. \ref{recieved distribution_small} (a).
	
	\item \textbf{Outer QAM/PSK constellation point (WR):} According to \eqref{QAM demodulation}, \eqref{PSK demodulation} and \eqref{discrete PDF}, when $s_k[l]=v_q$ and $v_q$ is the outer constellation point, ${\bar{y}}_k[l]$ with rescaling is expressed as 
	\begin{align}
	\overset{\qquad\qquad\qquad\qquad{\text{Gaussian random variable}}}{\underset{\text{discrete random variable located in }\mathcal{D}_k[l]\quad}{{\bar{y}}_k[l]\ \ =\ \ \underbrace{{\tilde s}_k[l]}\ \ +\ \ \overbrace{{n_k[l]}/{{\bar \gamma}}}}}.
	\label{distribution with rescaling}
	\end{align}
	The random variable ${\bar{y}}_k[l]$ is the sum of a discrete random variable ${\tilde s}_k[l]$ located in $\mathcal{D}_k[l]$ and an independent Gaussian random variable ${n_k[l]}/{{\bar \gamma}}$. The posteriori distribution of ${\tilde s}_k[l]$ is given in \eqref{discrete PDF}, and $f_{{ Y}}({{\bf y}}[l]|{\bf{v}}_q)$ is shown in Fig. \ref{recieved distribution_small}. It is worth noting that since only some combinations of ${\bf s}_{{\mathcal{K}}\backslash k}[l]$ can exploit CI from $s_k[l]$ and there still exists ${\tilde s}_k[l]=s_k[l]$ in most cases, which means $f_{{ Y}}({{\bf y}}[l]|{\bf{v}}_q)$ can be approximated as the PDF of the Gaussian distribution followed by $v_q+{n_k[l]}/{{\bar \gamma}}$ like \eqref{inner distribution with rescaling}.
	However, due to the presence of ${\tilde s}_k[l]$ extending towards CIR, assuming $f_{{ Y}}({{\bf y}}[l]|{\bf{v}}_q)$ to be Gaussian distributed will introduce a mismatching error for parameter estimation, i.e., 
	\begin{align}
	\begin{split}
		{\hat \sigma}^2_s &= \mathbb{E}\left\{({\bar{y}}_k[l]- s_k[l])^2\right\}\\
		&= \mathbb{E}\left\{({\tilde s}_k[l]-s_k[l])^2\right\} + \mathbb{E}\left\{({n_k[l]/{\bar \gamma}})^2\right\}\\
		&\geq \mathbb{E}\left\{({n_k[l]/{\bar \gamma}})^2\right\},
	\end{split}
	\label{sigma error}
	\end{align}
	where ${\hat \sigma}^2_s$ is the parameters that the Gaussian soft demodulator is expected to estimate in \eqref{Gaus PDF}, and $\mathbb{E}\left\{({n_k[l]/{\bar \gamma}})^2\right\}$ is the variance of the Gaussian distribution followed by $v_q+{n_k[l]}/{{\bar \gamma}}$. 
	
	\item \textbf{PSK constellation point (WOR):} According to \eqref{PSK demodulation} and \eqref{discrete PDF}, ${\bar{y}}_k[l]$ without rescaling is expressed as 
	\begin{align}
		\overset{\qquad\qquad\text{scalar random variable\quad \ \ Gaussian random variable}}{\underset{\qquad\text{discrete random variable located in }\mathcal{D}_k[l]}{{\bar{y}}_k[l]\ \ =\ \ \overbrace{{{\gamma[l]}}}\cdot\underbrace{{\tilde s}_k[l]}\ \ +\ \ \overbrace{n_k[l]}}}.
		\label{distribution without rescaling}
	\end{align}
	Different from \eqref{distribution with rescaling}, $\gamma[l]$ in \eqref{distribution without rescaling} can adjust the radial scaling of ${\tilde s}_k[l]$, resulting in ${\bar{y}}_k[l]$ exhibiting a distribution that resembles stripes, which is shown in Fig. \ref{recieved distribution_small} (c). 
	\end{itemize}
	
	In summary, ${f}_{{ Y}}({{\bf y}}|{\bf{v}}_q)$ with SLP usually does not follow Gaussian distribution. Thus, estimating the parameters of ${f}_{{ Y}}({{\bf y}}|{\bf{v}}_q)$ by treating it as Gaussian in the soft demodulator will lead to inaccurate LLRs, which will degrade the performance of the soft decoder.

	
	Including the analyzed CISB, different SLP schemes result in different distributions of the received signals. In order to facilitate the subsequent design of soft demodulators, we classify the received signals from different modulation and SLP schemes into the following two types:
	\begin{itemize}
	\item \textbf{Type I:} This type of signal consists of QAM received signals with SLP schemes like CISB and CIMMSE \cite{9910472}. The received signals of the inner QAM constellation points follow
	Gaussian distributions like Fig. \ref{recieved distribution_small} (a), while the received signals of the outer ones approximatly exhibit a Gaussian distribution with the exception of few probabilities extending along the CIR like Fig. \ref{recieved distribution_small} (b). We propose the soft demodulator for this type of signals in Section \ref{MGaus demodulator}.
	\item \textbf{Type II:} This type of signal consists of all PSK received signals (with and without rescaling). It also consists of QAM signals with SLP schemes whose signal distributions may deviate significantly from a Gaussian distribution \cite{shao2020minimum,9054488,8374931,9035662}. To provide an instance of such SLPs for the design and performance validation of the soft demodulator, we introduce the SLP for average SER minimization (ASM) in \cite{Wang2023}.\footnote{Its signals exhibit a non-Gaussian distribution, and for the sake of readability, we illustrate the distribution in Figure \ref{recieved distribution} (a3) and (b3) in Section \ref{results}.}  We propose soft demodulators for this type of signal in Sections \ref{GMM demodulator} and \ref{PFEN demodulator}.
	\end{itemize}
	
	\begin{figure*}[htp]
		\centering
		\includegraphics[width=7in]{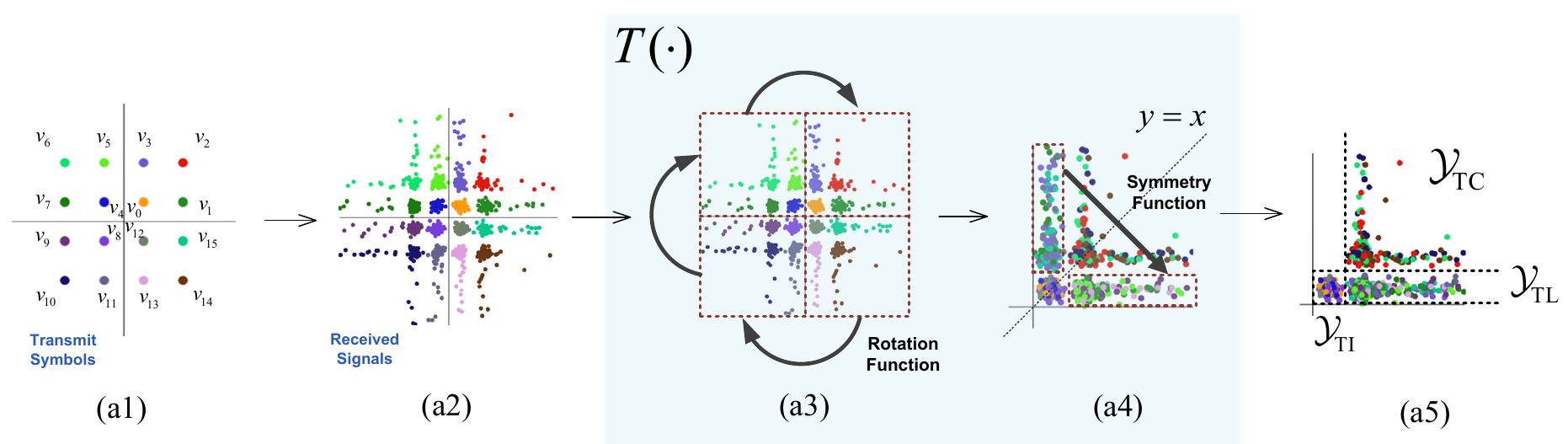}
		\caption{The constructions of ${{\mathcal{Y}}_{\rm TI}}$, ${{\mathcal{Y}}_{\rm TC}}$, and ${{\mathcal{Y}}_{\rm TL}}$ for 16QAM. The correspondence between the colors of received signals and the transmit symbols implies that the corresponding transmit symbols for each signal is known to the UE.}
		\label{fig: transorm function}
	\end{figure*}
	
	\section{Soft Demodulators for Non-Gaussian Received Signals}\label{All Soft Demodulators}
		\subsection{Modified Gaussian Soft Demodulator}\label{MGaus demodulator}
		In this subsection, we design the modified Gaussian soft demodulator for SLP schemes whose received signals belong to \textbf{Type I}. The soft demodulator approximates the received signal distribution of the outer QAM constellation point in \eqref{distribution with rescaling} as the Gaussian distribution with mean $v_q$. 
		As the Gaussian distribution in \eqref{distribution with rescaling} has the same variance as that in \eqref{inner distribution with rescaling}, the demodulator estimates the variance from the received signals belonging to inner QAM constellation points. It then applies the variance to $f_{{ Y}}({{\bf y}}[l]|{\bf{v}}_q)$ belonging to outer ones in \eqref{distribution with rescaling}, which prevents the problem of overestimating in \eqref{sigma error}.
		
		We consider pilot-assisted transmission, where a transmission block contains some pilot signals whose transmit symbols are known to UE. The received signals ${{\bf Y}}\in{\mathbb{R}}^{{ L}\times 2}$ in a transmission block at UE can be expressed as
		\begin{align}
			\begin{split}
				{{\bf Y}} &= \begin{bmatrix}
					{{\bf y}}[1] &{{\bf y}}[2] & \cdots & {{\bf y}}[L]
				\end{bmatrix}^T\\
				&=	\begin{bmatrix}
					{\bf Y}_{\rm p}^T & {\bf Y}_{\rm d}^T
				\end{bmatrix}^T,
			\end{split}
		\end{align}
		where ${\bf Y}_{\rm p}$ and  ${\bf Y}_{\rm d}$ are the received pilot signals and received data signals given by
		\begin{align}
			{\bf Y}_{\rm p} &= \begin{bmatrix}
				{\bf y}^{\rm p}[1] &{\bf y}^{\rm p}[2] & \cdots & {\bf y}^{\rm p}[{L_{\rm P}}]
			\end{bmatrix}^T\in{\mathbb{R}}^{{L_{\rm P}}\times 2},\\
			{\bf Y}_{\rm d} &= \begin{bmatrix}
				{\bf y}^{\rm d}[1] &{\bf y}^{\rm d}[2] & \cdots & {\bf y}^{\rm d}[{L_{\rm D}}]
			\end{bmatrix}^T\in{\mathbb{R}}^{{L_{\rm D}}\times 2},
		\end{align}
		where ${L_{\rm P}}$ is the number of pilot symbols, ${L_{\rm D}}$ is the number of data symbols, and ${L_{\rm P}}+{L_{\rm D}}=L$. According to the corresponding transmit symbols, the receiver can divide ${\bf Y}_{\rm p}$ into the following $Q$ signal sets:
		\begin{align}
			{\mathcal{Y}}_{0},{\mathcal{Y}}_{1},...,{\mathcal{Y}}_{q},...,{\mathcal{Y}}_{Q-1},
			\label{separate}
		\end{align}
		where ${\mathcal{Y}}_{q} = \{{\bf y}^{\rm p}_q[1],..,{\bf y}^{\rm p}_q[{L_{q}}]\}$ represents the set of received pilot signals whose transmit symbols are ${\bf v}_q$, and ${L_{q}}$ is the number of signals in ${\mathcal{Y}}_{q}$ which satisfies $\sum_{q=0}^{Q-1}{L_{q}}={L_{\rm P}}$. Note that signals in ${\mathcal{Y}}_{q}$ are the samples from $f_{{ Y}}({{\bf y}}|{\bf{v}}_q)$.

		For convenience in subsequent expressions, we define ${\mathcal{ Q}}=\{0, 1, ...,Q-1\}$ and denote the index sets of inner symbols, corner symbols, and lateral symbols as $\mathcal{Q}_{\rm I}$, $\mathcal{Q}_{\rm C}$, and $\mathcal{Q}_{\rm L}$, respectively. Besides, we set the symbol indexes ${\mathcal{Q}}$ in consecutive order on the constellation map, based on which $\mathcal{Q}_{\rm I}$, $\mathcal{Q}_{\rm C}$ and $\mathcal{Q}_{\rm L}$ can be determined. For example, the symbol indexes for 16QAM is shown in Fig. \ref{fig: transorm function} (a1), and we have
		\begin{align}
			\begin{split}
				&\mathcal{Q}_{\rm I}=\{0, 4, 8, {12}\}, \mathcal{Q}_{\rm C}=\{2, 6, {10}, {14}\},\\
				&\quad \quad\mathcal{Q}_{\rm L}=\{1,3,5,7,9,11,13,15\}.
			\end{split}
		\end{align}
		For PSK, we define $v_q=\exp[j(q+\frac{1}{2})\frac{2\pi}{Q}],\forall q\in{{\mathcal{ Q}}}$, and $\mathcal{Q}_{\rm C}={\mathcal{ Q}}$.
		
		The Gaussian variance is estimated from the received signals belonging to inner symbols, i.e., 
		\begin{align}
			{\hat\sigma}^2_{\rm is} = \frac{1}{\sum\limits_{q\in{\mathcal{Q}_{\rm I}}}L_{q}}\sum_{q\in{\mathcal{Q}_{\rm I}}}\sum^{{L_{q}}}_{l=1}\|{\bf y}^{\rm p}_q[l]-{\bf v}_q\|^2_2.
			\label{inner variance estimation}
		\end{align}
		The likelihood functions in the modified Gaussian soft demodulator have the same formulation as the conventional soft demodulator \eqref{Gaus PDF} while replacing ${\hat\sigma}^2_{\rm s}$ with ${\hat\sigma}^2_{\rm is}$.
		
	\subsection{GMM Soft Demodulator}\label{GMM demodulator}
	The modified Gaussian soft demodulator still faces challenges in computing LLRs for received signals belonging to \textbf{Type II}, where the distribution of QAM signals or PSK received signals without rescaling deviates significantly from a Gaussian distribution. Although the distribution of rescaled PSK received signals may be approximated by a Gaussian distribution, such as the received signals in \eqref{distribution with rescaling} from CISB and CIMMSE, the modified Gaussian soft demodulator still cannot be applied due to $\mathcal{Q}_{\rm I}=\emptyset$ in PSK. To address this issue, we design the GMM soft demodulator, which employs GMM to approximate $f_{{Y}}({\bf y}|{\bf{v}}_q)$ from received pilot signals. 
	
	GMM is a mixture of several Gaussian distributions, and its PDF for 2D samples is given by \cite{reynolds2009gaussian} 
	\begin{align}
		f_{\rm g}({\bf y};\mathcal{P}) =
		\sum_{n=1}^{N_{\rm g}}a_n\frac{\exp\left[{-\frac{({\bf y}-\boldsymbol{\mu}_n)^T\boldsymbol{\Sigma}^{-1}({\bf y}-\boldsymbol{\mu}_n)}{2}}\right]}{2\pi\cdot\det(\boldsymbol{\Sigma}_n)^{\frac{1}{2}}},
		\label{E3}
	\end{align}
	where $N_{\rm g}$ denotes the order of GMM. The parameters of the GMM are given by $\mathcal{P}=\left\{\boldsymbol{\mu}_{n}, \boldsymbol{\Sigma}_{n}, a_{n}\right\}^{N_{\rm g}}_{n=1}$, where ${\boldsymbol {\mu}}_n\in\mathbb{ R}^{2\times 1}$, $\boldsymbol{\Sigma}_n\in\mathbb{ R}^{2\times 2}$, and $a_n\in\mathbb{ R}$ denote expectation, covariance, and weight of $n$-th Gaussian distribution, respectively. GMM can be used to approximate a certain distribution, i.e., there exist $Q$ GMMs satisfying
	\begin{align}
		f_{{ Y}}({{\bf y}}|{\bf{v}}_q) \approx f_{\rm g}({{\bf y}} ; \mathcal{P}_{q}),\;\forall q \in {\mathcal{ Q}},
		\label{approx1}
	\end{align}
	where $\mathcal{P}_{q}$ denotes the GMM parameters for $f_{{ Y}}({{\bf y}}|{\bf{v}}_q)$. Thus, PDF ${{{f}_{{ Y}}}}({\bf y}|{\bf{v}}_q)$ can be obtained by finding the best-matched $\mathcal{P}_{q}$. As the signals in ${\mathcal{Y}}_{q}$ are discrete samples of $f_{{ Y}}({\bf y}|{\bf{v}}_q)$, $\mathcal{P}_{q}$ can be estimated by certain estimation algorithm $C(\cdot)$ as \cite{kay1993fundamentals}
	\begin{align}
		\begin{split}
			{\widehat{\mathcal{P}}}_{q}=C\left({\mathcal{Y}}_{q}\right) ,\;\forall q \in {\mathcal{ Q}}.
		\end{split}
		\label{GMM0}
	\end{align}
	
	Since the number of unknown parameters is increasing with the number of PDFs to be estimated, it requires a larger $L_q$ to obtain accurate estimates of these unknown parameters, which introduces a significant pilot overhead. Thus, how to reduce the number of PDFs to be estimated by revealing the correlation between received symbols is a crucial issue in GMM soft demodulators. 
	
	\begin{ppn}
	For PSK and 16QAM transmission with SLP schemes CISB, CIMMSE or ASM, $f_{{ Y}}({{\bf y}}|{\bf{v}}_m)$ and $f_{{ Y}}({{\bf y}}|{\bf{v}}_n)$ are related as follows
		\begin{align}
			{f}_{{ Y}}\left({\bf y}|{\bf{v}}_m\right) = {f}_{{ Y}}\left(R_{m-n}({\bf y})|{\bf{v}}_n\right),\ m,n\in {\mathcal{ Q}},
			\label{shift P}
		\end{align}	
		where $m$ and $n$ should satisfy $\frac{m-n}{4}\in{\mathbb{Z}}$ for 16QAM transmission. $R_{t}(\cdot),t\in {\mathbb{N}}$ is the rotation functon defined as
		\begin{align}
			R_{t}\left(\begin{bmatrix}
				\real\left({\bar y}\right)\\
				\imaginary\left({\bar y}\right)
			\end{bmatrix}\right)=\begin{bmatrix}
				\real\left({\bar y}\cdot\exp(-jt\frac{2\pi}{Q})\right)\\
				\imaginary\left({\bar y}\cdot\exp(-jt\frac{2\pi}{Q})\right)
			\end{bmatrix}.
		\end{align}
		\label{rotation ppn}
	\end{ppn}
	\begin{pf}
	See Appendix \ref{ppn 1 proof}.
	\end{pf}
	
	\begin{remark}
	The above proposition indicates that the received signal density of ${\bf v}_m$ can be rotated into that of ${\bf v}_n$. Following this proposition, 
	$\{f_{{ Y}}({{\bf y}}|{\bf{v}}_q)\}_{q\in{\mathcal{ Q}}}$ of PSK and QAM transmission can be respectively obtained from $f_{{ Y}}({{\bf y}}|{\bf{v}}_0)$ and $\{f_{{ Y}}({{\bf y}}|{\bf{v}}_q)\}_{q\in\{0,1,2,3\}}$ by utilizing $R_{t}(\cdot)$, which significantly reduces the number of parameters to be estimated.
	\end{remark}
	
	For 16QAM transmissioin, we further make the approximation that the received signal distributions of ${\bf v}_1$ and ${\bf v}_3$ are symmetric about the line $y=x$, i.e., 
	\begin{align}
		{f}_{{ Y}}\left({\bf y}|{\bf{v}}_1\right) = {f}_{{ Y}}\left(S({\bf y})|{\bf{v}}_3\right),
		\label{symmetry}
	\end{align}
	where $S(\cdot)$ is the symmetry functon defined as
	\begin{align}
		S\left(\begin{bmatrix}
			\real\left({\bar y}\right)\\
			\imaginary\left({\bar y}\right)
		\end{bmatrix}\right)=\begin{bmatrix}
			\imaginary\left({\bar y}\right)\\
			\real\left({\bar y}\right)
		\end{bmatrix}.
	\end{align}
	Therefore, only $\{f_{{ Y}}({{\bf y}}|{\bf{v}}_q)\}_{q\in\{0,1,2\}}$ requires to be estimated. Since ${f}_{{ Y}}\left({\bf y}|{\bf{v}}_0\right)$, ${f}_{{ Y}}\left({\bf y}|{\bf{v}}_1\right)$, and ${f}_{{ Y}}\left({\bf y}|{\bf{v}}_2\right)$ respectively contain the distribution properties of the received signals belonging to the inner, corner, and lateral constellation points, we use $f_{Y_{\rm I}}({\bf y})$, $f_{Y_{\rm C}}({\bf y})$, and $f_{Y_{\rm L}}({\bf y})$ to denote them. Combining \ppnref{rotation ppn} and \eqref{symmetry}, we have
	\begin{align}
		{{{f}_{{ Y}}}}({\bf y}|{\bf{v}}_q)=  
		\begin{cases}
			f_{Y_{\rm I}}(T\left({\bf y}, q\right)),\ {\rm if}\ q \in {\mathcal{ Q}}_{\rm I}\\
			f_{Y_{\rm C}}(T\left({\bf y}, q\right)),\ {\rm if}\ q \in {\mathcal{ Q}}_{\rm C}\\
			f_{Y_{\rm L}}(T\left({\bf y}, q\right)),\ {\rm if}\ q \in {\mathcal{ Q}}_{\rm L}
		\end{cases},
		\label{I D L}
	\end{align}
	where $T\left(\cdot\right)$ is the transform function given by
	\begin{align}
		\begin{split}
			T\left({\bf y}, q\right) = \begin{cases}
				S(R_{4\lfloor\frac{q}{4}\rfloor}({\bf y})),\ {\rm if}\ q\in\{3,7,11,15\}\\
				R_{4\lfloor\frac{q}{4}\rfloor}({\bf y}),\ {\rm otherwise}
			\end{cases}.
		\end{split}
	\end{align}
	
	
	Given certain sample ${\bf y}^{\rm p}_{m}[l]$ from ${f}_{{ Y}}\left({\bf y}|{\bf{v}}_m\right)$, $T({\bf y}^{\rm p}_{m}[l])$ can be seen as the sample from ${f}_{{ Y}}\left({\bf y}|{\bf{v}}_n\right)$ if the received signal distribution of ${\bf{v}}_m$ can be transfomed into that of ${\bf{v}}_n$ by $T(\cdot)$, i.e., ${f}_{{ Y}}\left({\bf y}|{\bf{v}}_m\right)={f}_{{ Y}}\left(T({\bf y})|{\bf{v}}_n\right)$. Thus, we define 
	\begin{align}
			{{\mathcal{Y}}_{\rm TI}} = \left\{T\left({\bf y}^{\rm p}_q, q\right)\bigg|{\bf y}^{\rm p}_{q}\in \bigcup\limits_{q'\in {\mathcal{ Q}}_{\rm I}} {\mathcal{Y}}_{q'}\right\},
		\label{Y_I,D,L}
	\end{align}
	where ${{\mathcal{Y}}_{\rm TC}}$ and ${{\mathcal{Y}}_{\rm TL}}$ can be obtained in a similar manner by replacing ${\mathcal{ Q}}_{\rm I}$ with ${\mathcal{ Q}}_{\rm C}$ and ${\mathcal{ Q}}_{\rm L}$. ${{\mathcal{Y}}_{\rm TI}}$, ${{\mathcal{Y}}_{\rm TC}}$, and ${{\mathcal{Y}}_{\rm TL}}$ can be seen as the samples from $f_{Y_{\rm I}}({\bf y})$, $f_{Y_{\rm C}}({\bf y})$, and $f_{Y_{\rm L}}({\bf y})$ respectively. The construction of these three sets is shown in Fig. \ref{fig: transorm function} (a1)-(a5).  Then, we use GMM to approximate the three probability density functions. We illustrate the process with $f_{{ Y_{\rm I}}}({\bf y})$ as an instance:
	\begin{align}
			f_{{ Y_{\rm I}}}({\bf y}) \approx f_{\rm g}({\bf y} ; \mathcal{P}_{\rm I}) &,\ {\widehat{\mathcal{P}}}_{\rm I}=C\left({\mathcal{Y}}_{\rm TI}\right).
		\label{3 GMM}
	\end{align}
    
\begin{figure}[t]
		\centering
		\includegraphics[width=3in]{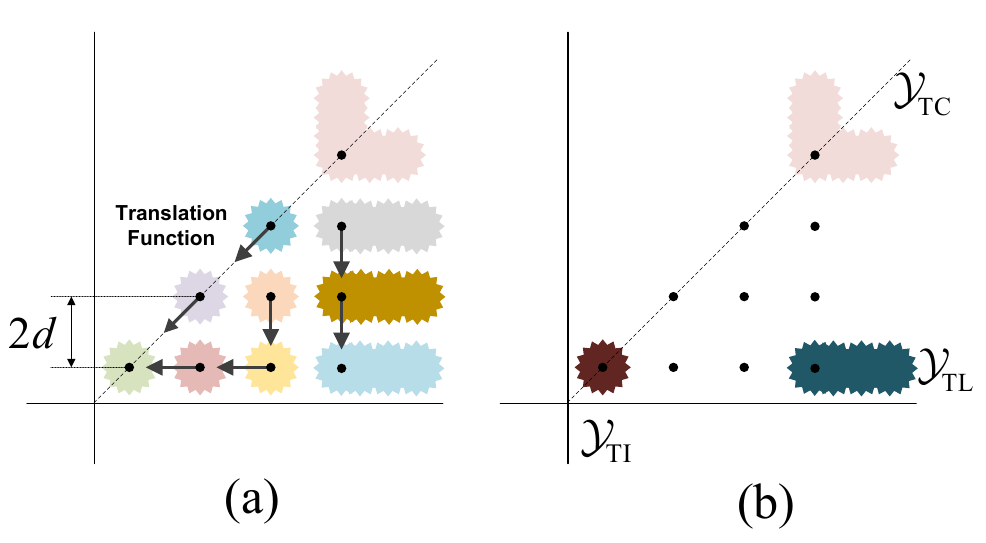}
		\caption{The constructions of ${{\mathcal{Y}}_{\rm TI}}$, ${{\mathcal{Y}}_{\rm TC}}$, and ${{\mathcal{Y}}_{\rm TL}}$ for 64QAM (The process of Fig.\ref{fig: transorm function} (a1)-(a4) is omitted). $d$ is half of the distance between adjacent constellation points.}
		\label{translation_function}
	\end{figure}

	Due to the increased number of inner and lateral constellation points in higher-order QAM, $R_t(\cdot)$ and $S(\cdot)$ alone are insufficient. Therefore, we further introduce the translation function depicted in Fig. \ref{translation_function} for $T(\cdot)$. Fig. \ref{translation_function} (a) shows the signal distributions after process of Fig.\ref{fig: transorm function} (a1)-(a4). Received signals are then  translated based on the relative positions of the corresponding transmit symbols so as to merge the received signals belonging to the inner and lateral symbols,  respectively, as shown in Fig. \ref{translation_function} (b). We omit the mathematical definition of the translation function due to space limitations. 
	
	For PSK transmission, formulation \eqref{I D L} and \eqref{Y_I,D,L} are also capable, where ${\mathcal{ Q}}_{\rm I}={\mathcal{ Q}}_{\rm L}=\emptyset$. $f_{Y_{\rm C}}$ denotes ${f}_{{ Y}}\left({\bf y}|{\bf{v}}_0\right)$ and 
	\begin{align}
		T\left(
		{\bf y}, q\right) = 
		R_{q}({\bf y}),\ \forall q \in {\mathcal{ Q}}.
		\label{psk T}
	\end{align}	
	
	For the same ${L_{\rm P}}$, the performance of $C(\cdot)$ in \eqref{3 GMM} is much better than that in \eqref{GMM0} since the numbers of samples in ${{\mathcal{Y}}_{\rm TI}}$, ${{\mathcal{Y}}_{\rm TC}}$, and ${{\mathcal{Y}}_{\rm TL}}$ are $|\mathcal{Q}_{\rm I}|$, $|\mathcal{Q}_{\rm C}|$, and $|\mathcal{Q}_{\rm L}|$ times greater than that of ${\mathcal{Y}}_{q}$ in \eqref{GMM0}. By defining ${{\hat{f}_{{ Y}}}}({\bf y}^{\rm d}[l]|{\bf{v}}_q)$ and ${\widehat{\rm LLR}}_i[l]$ as the estimation of ${{{f}_{{ Y}}}}({\bf y}^{\rm d}[l]|{\bf{v}}_q)$ and the LLR belonging to $i$-th bit of ${\bf y}^{\rm d}[l]$, the GMM soft demodulator is presented in Algorithm \ref{A3}, which is also capable for PSK by setting  $\mathcal{Q}_{\rm I}=\mathcal{Q}_{\rm L}={\mathcal{Y}}_{\rm TI}={\mathcal{Y}}_{\rm TL}=\emptyset$.
	

	\begin{table*}[htb]
		\renewcommand\arraystretch{1.3}
		\scriptsize
		\centering
		\caption{Network Structure and Major Parameters of SubModel for ${{\mathcal{Y}}_{I}}$}
		\label{table-submodel}
		
		\resizebox{1.2\columnwidth}{!}
		{\begin{tabular}{c c c c|c|c c}\toprule
				\multicolumn{4}{c|}{\textbf{MAB}} &  \textbf{FC Size} &\multicolumn{2}{|c}{\textbf{Matrix Size}}\\ \midrule 
				&MAB1&MAB2& MAB2 & FFC & ${\bf E}_1$ & ${\bf E}_2$               \\
				
				Heads Number& 4 & 4 & 4  & $32\times 6$ & $16\times 32$ & $N_{\rm g}\times 32$ 
				\\
				FC1 Size& $32\times 32$ & $4\times 32$& $32\times 32$  &  & ${\bf M}_{\rm I}$  &  ${\bf O}_{\rm I}$ 
				\\
				FC2/3 Size& $4\times 32$ & $32\times 32$  & $32\times 32$ &  &  ${{L}_{\rm I}}\times 3$ &  $N_{\rm g}\times 6$
				\\
				FC4 Size& $32 \times 32$& $32\times 32$ & $32\times 32$ &  &   &   
				\\
				\bottomrule 
		\end{tabular}}
	\end{table*}
	
	\begin{algorithm}[t]
		\caption{GMM Soft Demodulator}
		\label{A3}
		\begin{spacing}{1.0}
			\begin{algorithmic}[1]
				\STATE \textbf{Input: }$\mathcal{Q}_{\rm I}$, $\mathcal{Q}_{\rm L}$, $\mathcal{Q}_{\rm C}$,  $\{{\mathcal{Y}}_{q}\}_{q\in {\mathcal{ Q}}}$,  ${\bf Y}_{\rm d}$.
				\STATE Get ${{\mathcal{Y}}_{\rm TI}}$, ${{\mathcal{Y}}_{\rm TC}}$, and ${{\mathcal{Y}}_{\rm TL}}$ using \eqref{Y_I,D,L}.
				\STATE Estimate GMM parameters:\\
				${\widehat{\mathcal{P}}}_{\rm I}=C({\mathcal{Y}}_{\rm TI})$, ${\widehat{\mathcal{P}}}_{\rm C}=C({\mathcal{Y}}_{\rm TC})$, ${\widehat{\mathcal{P}}}_{\rm L}=C({\mathcal{Y}}_{\rm TL})$.
				\STATE \textbf{for} $l=1$ \textbf{to} $L_{\rm D}$ \textbf{do} \label{llr start}
				\STATE \quad \textbf{for} $q=0$ \textbf{to} $Q-1$ \textbf{do} 
				
				\STATE \quad \quad ${{{\hat f}_{{Y}}}}({\bf y}^{\rm d}[l]|{\bf{v}}_q)=\begin{cases}
					f_{\rm g}(T\left({\bf y}^{\rm d}[l], q\right)| {\widehat{\mathcal{P}}}_{\rm I})\ \ {\rm if}\   q \in {\mathcal{ Q}}_{\rm I}\\
					f_{\rm g}(T\left({\bf y}^{\rm d}[l], q\right)| {\widehat{\mathcal{P}}}_{\rm C})\ \ {\rm if}\   q \in {\mathcal{ Q}}_{\rm C}\\
					f_{\rm g}(T\left({\bf y}^{\rm d}[l], q\right)| {\widehat{\mathcal{P}}}_{\rm L})\ \ {\rm if}\   q \in {\mathcal{ Q}}_{\rm L}\\
				\end{cases}$.
				\STATE \quad \textbf{end} \textbf{for} 
				\STATE \quad  \textbf{for} $i=1$ \textbf{to} ${\log_2Q}$ \textbf{do} 
				\STATE \quad \quad ${\widehat{\rm LLR}}_i[l]= {\rm ln}\frac{\sum\limits_{{\bf{v}}_q\in{{\mathcal{S}}^{+}_i}}{{{\hat f}_{{Y}}}}({\bf y}^{\rm d}[l]|{\bf{v}}_q)}{\sum\limits_{{\bf{v}}_q\in{{\mathcal{S}}^{-}_i}}{{{\hat f}_{{Y}}}}({\bf y}^{\rm d}[l]|{\bf{v}}_q)}$.
				\STATE \quad \textbf{end} \textbf{for} 
				\STATE \textbf{end} \textbf{for} \label{llr end}
				\STATE \textbf{Output:}  $\left\{{\widehat{\rm LLR}}_i[l]\right\}^{\log_2{Q}}_{i=1}$, $\forall l =1,...,{L_{\rm D}}$.
			\end{algorithmic}\label{GMM algorithm}
		\end{spacing}
	\end{algorithm}
	
	\subsection{Pilot Feature Extraction Network}\label{PFEN demodulator} 
\begin{figure}[hbtp]
		\centering
		\includegraphics[width=3.5in]{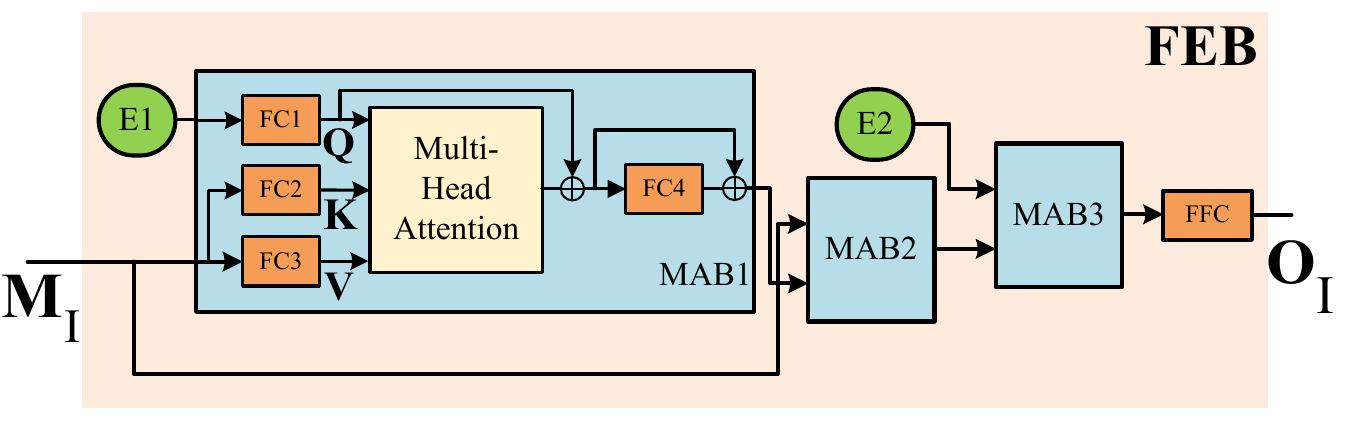}
		\caption{Structure of FEB.}
		\label{PFEN}
	\end{figure}

	To further reduce the pilot overhead, we propose PFEN, a low-complexity demodulator that utilizes the feature extraction block (FEB) to extract the signal distribution features from ${{\mathcal{Y}}_{\rm TI}}$, ${{\mathcal{Y}}_{\rm TC}}$, and ${{\mathcal{Y}}_{\rm TL}}$.
	Since elements in these three sets are order-independent, FEB is expected to be a permutation-invariant function \cite{NIPS2017_f22e4747}, which can be achieved by the outstanding transformer-based operations \cite{lee2019set}. The structure of PFEN is shown in Fig. \ref{PFEN}, where three FEBs are used to process ${{\mathcal{Y}}_{\rm TI}}$, ${{\mathcal{Y}}_{\rm TC}}$, and ${{\mathcal{Y}}_{\rm TL}}$ individually. Note that the transformer-based operations are also well-suitable for pilot signals with adjustable lengths in practical communication systems since their pre-trained model can deal with varying numbers of inputs \cite{9298921}.
	
	\subsubsection{FEB} Since the structure of three FEBs for ${{\mathcal{Y}}_{\rm TI}}$, ${{\mathcal{Y}}_{\rm TC}}$, and ${{\mathcal{Y}}_{\rm TL}}$ are the same, we only introduce the one for ${{\mathcal{Y}}_{\rm TI}}$. The parameters of the FEB for ${{\mathcal{Y}}_{\rm TI}}$ are listed in Table \ref{table-submodel}: `Heads Number' represents the number of heads of `Multi Head Attention' module \cite{vaswani2017attention};
	`FC' and `FFC' denote fully connected layer applyied at the last dimension; ${\bf E}_1$ and ${\bf E}_2$ are trainable matrices introduced for reducing the dimension of variables involved in computation \cite{lee2019set}; ${\bf M}_{\rm I}$ and ${\bf O}_{\rm I}$ are the input and output. Except for the received pilot signals, we also choose ${\bar \gamma}$ as the input to provide more information for PFEN.
	The input matrix ${\bf M}_{I}$ is given by
	\begin{align}
		\begin{split}
			{\bf M}_{\rm I} &= \left[{\bf m}[1]^T,{\bf m}[2]^T,...,{\bf m}[{{L}_{\rm I}}]^T\right]^T \in \mathbb{R}^{{{L}_{\rm I}}\times 3},\\
			{\bf m}[l] &= \left[{{\bf y}^{\rm p}_{\rm TI}}[l]^T, {\bar \gamma}\right] \in \mathbb{R}^{1\times 3},\ l=1,...,{{L}_{\rm I}},
		\end{split}
		\label{input}
	\end{align}
	where ${{\bf y}^{\rm p}_{\rm TI}}[l]$ denotes $l$-th signal in ${{\mathcal{Y}}_{\rm TI}}$, and $L_{\rm I}=|{{\mathcal{Y}}_{\rm TI}}|$.
	
	The parameters in the $n$-th Gaussian distribution can be expressed as 
	\begin{align}
		\alpha_n, {\boldsymbol{\mu}} = 
		\begin{bmatrix}
			\mu_{n,1}\\
			\mu_{n,2}
		\end{bmatrix}
		, \boldsymbol{\Sigma}_{n} = 
		\begin{bmatrix}
			\sigma^2_{n,1} & \sigma^2_{n,2}\\
			\sigma^2_{n,2} & \sigma^2_{n,3}
		\end{bmatrix},
	\end{align}
	which contains 6 parameters. Denote the output as ${\bf O}_{\rm I}\in \mathbb{R}^{N_{\rm g}\times 6}$, and the process of FEB can be expressed as
	\begin{align}
		{\bf O}_{\rm I} = {\rm FEB}_{\rm I}\left({\bf M}_{\rm I}\right).
	\end{align}
	Due to the permutation invariance of FEB, rows of ${\bf O}_{\rm I}$ and ${\bf M}_{\rm I}$ are unordered, which means ${\bf O}_{\rm I}$ and ${\bf M}_{\rm I}$ can be regarded as two sets with their rows as elements.
	
	\subsubsection{GPL}We design a GMM property layer (GPL) to ensure that the outputs meet the required properties of GMM parameters, i.e., 
	\begin{align}
		{\widehat{\mathcal{P}}}_{\rm I} = {\rm GPL}\left({\bf O}_{\rm I}\right).
	\end{align}
	GPL performs the following operations:
	\begin{itemize}
		\item Weight: To keep $\sum_{n=1}^{N_{\rm g}}a_{n}=1$, GPL applys
		\begin{align}
			{\hat a}_1, ..., {\hat a}_{N_{\rm g}} = {\rm Softmax}\left(o_{1, 1}, ..., o_{N_{\rm g}, 1}\right),
		\end{align}
		where $o_{i,j}$ is the $(i,j)$-th element of ${\bf O}_{\rm I}$.
		\item Mean: ${\hat{\boldsymbol{\mu}}}_n=[o_{n, 2},o_{n, 3}]^T,\ n=1,...,N_{\rm g}$.
		\item Covariance: To keep ${\hat{\boldsymbol{\Sigma}}}_{n}$ positive-definite and its diagonal elements non-negative, GPL applies
		\begin{align}
			\begin{split}
				&{\hat \sigma}^2_{n,1} = {\rm SoftPlus}\left(o_{n,4}\right),\ {\hat \sigma}^2_{n,3} = {\rm SoftPlus}\left(o_{n,6}\right),\\
				&{\hat \sigma}^2_{n,2}  = \sqrt{{\hat \sigma}^2_{n,1}{\hat \sigma}^2_{n,3}}\cdot {\rm Tanh}\left(o_{n,5}\right),\ n=1,...,N_{\rm g}.
			\end{split}
		\end{align}
	\end{itemize}
	
	\subsubsection{Loss Function}To facilitate the computation of the loss function during training, we also separate ${\bf Y}_{\rm d}$ according to \eqref{separate} and apply $T\left(\cdot\right)$ to transform the separated sets into ${\mathcal{Y}}^{\rm d}_{\rm TI}$, ${\mathcal{Y}}^{\rm d}_{\rm TC}$, and ${\mathcal{Y}}^{\rm d}_{\rm TL}$ like \eqref{Y_I,D,L}.
	Similar but not identical to the ML criterion that tries to maximize the log-likelihood function of samples, we hope the parameter extracted from ${{\mathcal{Y}}_{\rm TI}}$ could maximize the likelihood of ${\mathcal{Y}}^{\rm d}_{\rm TI}$. This idea ensures parameters extracted from the received pilot signals are optimized to match the received data signals, and the loss function is designed as 
	\begin{align}
		\begin{split}
			&{\rm Loss}_{\rm I} = \\
			&-{\mathbb{E}}_{{\mathcal{Y}}^{\rm d}_{\rm TI}}\left[\ln\sum_{n=1}^{N_{\rm g}}{\hat \alpha}_{n}\frac{\exp\left[{-\frac{\left({\bf y}^{\rm d}_{\rm TI}- {\hat{\boldsymbol{\mu}}}_n\right)^{T}{\hat{\boldsymbol{\Sigma}}}^{-1}_{n}\left({\bf y}^{\rm d}_{\rm TI}- {\hat{\boldsymbol{\mu}}}_n\right)}{2}}\right]}{2\pi\det({\hat{\boldsymbol{\Sigma}}}_{n})^{\frac{1}{2}}}  
			\right],
		\end{split}
	\end{align}
	where ${\bf y}^{\rm d}_{\rm TI}$ is the sample in ${\mathcal{Y}}^{\rm d}_{\rm TI}$. 
	
	PFEN for PSK has a similar structure and training process to those for QAM. Their difference is that PFEN for PSK only needs a FEB due to ${\mathcal{Y}}_{\rm TI}={\mathcal{Y}}_{\rm TL}=\emptyset$. For the received signals without power allocation and rescaling, ${\bar \gamma}$ is removed from the input \eqref{input}.
	
	\begin{figure*}[htp]
		\centering
		\includegraphics[width=7in]{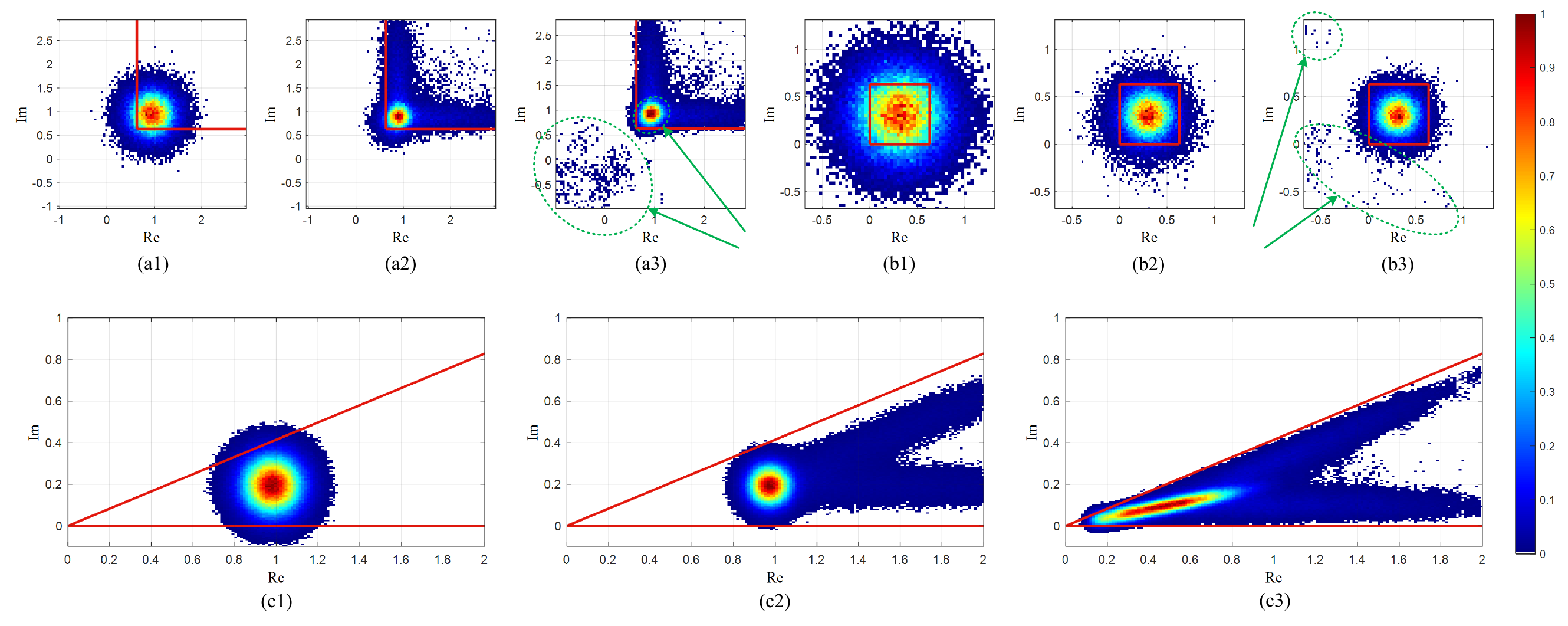}
		\caption{Normalized $f_{{ Y}}({{\bf y}}[l]|{\bf{v}}_q)$ where ${\bf{v}}_q$ is\\
		(a1) corner 16QAM constellation point, ZF. (a2) corner 16QAM constellation point, CIMMSE. (a3) corner 16QAM constellation point, ASM. 
		\\ (b1) inner 16QAM constellation point, ZF. (b2) inner 16QAM constellation point, CIMMSE. (b3) inner 16QAM constellation point, ASM. 
		\\(c1) 16PSK constellation point, ZF, WR. (c2) 16PSK constellation point, CIMMSE, WR. (c3) 16PSK constellation point, CIMMSE, WOR.\\  $N=K=8$, ${\rm SNR}=\frac{P_{\rm T}}{\sigma^2}$ is 20dB, the red lines represent the decision boundaries of the constellation points.}
		\label{recieved distribution}
	\end{figure*}
	
	\begin{algorithm}[t]
		\caption{PFEN Soft Demodulator}
		\begin{spacing}{1.1}
			\begin{algorithmic}[1]
				\STATE \textbf{Input: }$\mathcal{Q}_{\rm I}$, $\mathcal{Q}_{\rm L}$, $\mathcal{Q}_{\rm C}$,  $\{{\mathcal{Y}}_{q}\}_{q\in {\mathcal{ Q}}}$,  ${\bf Y}_{\rm d}$,${\bar \gamma}$.
				\STATE Get ${{\mathcal{Y}}_{\rm TI}}$, ${{\mathcal{Y}}_{\rm TC}}$, and ${{\mathcal{Y}}_{\rm TL}}$ using \eqref{Y_I,D,L}.
				\STATE Get ${\bf M}_I$, ${\bf M}_C$, and ${\bf M}_L$ using \eqref{input}.
				\STATE 
				${\bf O}_{\rm I} = {\rm FEB}_{\rm I}\left({\bf M}_{\rm I}\right)$, ${\widehat{\mathcal{P}}}_{\rm I} = {\rm GPL}\left({\bf O}_{\rm I}\right)$.
				\STATE ${\bf O}_{\rm C} = {\rm FEB}_{\rm C}\left({\bf M}_{\rm C}\right)$, ${\widehat{\mathcal{P}}}_{\rm C} = {\rm GPL}\left({\bf O}_{\rm C}\right)$.
				\STATE ${\bf O}_{\rm L} = {\rm FEB}_{\rm L}\left({\bf M}_{\rm L}\right)$, ${\widehat{\mathcal{P}}}_{\rm L} = {\rm GPL}\left({\bf O}_{\rm L}\right)$.
				\STATE The same as the steps 4-11 in Algorithm \ref{GMM algorithm}.
				\STATE \textbf{Output:}  $\left\{{\widehat{\rm LLR}}_i[l]\right\}^{\log_2{Q}}_{i=1}$, $\forall l =1,...,{L_{\rm D}}$.
			\end{algorithmic}\label{PFEN algorithm}
		\end{spacing}
	\end{algorithm}
	
	\section{Numerical Results}\label{results}
	In this section, we use the Monte Carlo method to evaluate the performance of the proposed methods in the scenario of an MU-MISO system and Rayleigh fading channel. Unless otherwise specified, we set $P_{\rm T}[l]=1$ and ${\rm SNR}=\frac{1}{\sigma^2}$. We consider the conventional ZF precoding scheme and three SLP schemes, CISB \cite{8466792,Li2021}, CIMMSE \cite{9910472}, and ASM \cite{Wang2023}. The proposed demodulators are compared with the Gaussian demodulator and the demodulator based on the Class A model in \cite{9645175}, denoted as `CA'. For CA, the estimation method of power parameter $\sigma^2_A$ is the same as the Gaussian demodulator, and the term number is set to $M_{\rm ca}=20$ \cite{9645175}.
	
	\subsection{Training Details and Complexity Analysis}
	\subsubsection{Training Details}
	The training details of PFEN are as follows. To improve the diversity of the dataset, we record ${\bf h}^T{\bf x}[l]$ instead of ${\bf h}^T{\bf x}[l]+n[l]$ and generate noise randomly at the training phase. The size of the training dataset for ${\bf h}^T{\bf x}[l]$ is $[300, N_{\rm SNR}, K, L_{\rm P}+L_{\rm D}, 2]$, which contains 300 channels, the scenarios of $N_{\rm SNR}$ different SNR values, and $K$ UEs. It should be noted that PFEN is trained to work in full SNR ranges, and the training data is not used for performance evaluation. For each channel, SNR, and UE, the received signals are stored as real and imaginary parts, and
	the first $L_{\rm P}=1024$ or $128$ (resp. QAM or PSK) signals are ${\bf Y}_{\rm p}$, and the latter $L_{\rm D}=2048$ signals are ${\bf Y}_{\rm d}$. For the convenience of batch computing, we set ${L_{q}}={L_{\rm P}}/Q,\ \forall q \in {\mathcal{ Q}}$ when training, while they can be unequal for validation since the amount of input samples is adjustable for a pre-trained transformer module. Similarly, ${L_{\rm P}}$ can be flexibly adjusted, and $L_{\rm D}$ could be expanded to be much larger when PFEN is employed in practical systems. We adopt the Adam optimizer with the learning rate progressively decreasing from $10^{-4}$ to $10^{-6}$, respectively. The number of iterations and batch size are set to be 90000 and 32.
	
	\subsubsection{Computational Complexity of GMM Soft Demodulator}
	\begin{table*}[hb]
	\renewcommand\arraystretch{1.3}
	\centering
	\caption{Computational Complexities of Soft Demodulators}
	\resizebox{2\columnwidth}{!}
	{
	\begin{tabular}{c|ccc|ccc|cc}
	\toprule[1.0pt]
   & \multicolumn{3}{c|}{\textbf{QAM}} & \multicolumn{3}{c|}{\textbf{PSK}} &   \textbf{Demodulator}   & \textbf{Complexity Order} \\ \midrule
   \textbf{Demodulator} & CA    & GMM    & PFEN    & CA    & GMM    & PFEN    & CA   & $\mathcal{O}\left(\eta_{\rm ca}{L_{\rm P}}M_{\rm ca}\right)$              \\
	\textbf{Number of Multiplications ($10^5$)} & $677.64$     & $116.85$      & $37.85$       & $80.48$     & $8.75$      & $5.45$       & GMM  & $\mathcal{O}\left(\eta_{\rm g}{L_{\rm P}}N_{\rm g}\right)$              \\
	\textbf{Execution Time (ms)}       & $539.52$     & $82.27$  & $5.76$       & $132.01$     & $9.80$      & $1.69$       & PFEN & $\mathcal{O}\left({L_{\rm P}}D_{\rm SA}\left({D_{{\bf M}}}+N_{{\bf E}_1}\right)+D^2_{\rm SA}N_{{\bf E}_1}\right)$              \\ \bottomrule[1.0pt]
	\end{tabular}
	}
	\label{complexity comparison}
	\end{table*}
	Since the prior information about the signal distribution is unknown, we choose the ML criterion instead of the MAP criterion for parameter estimation \cite{zhang2022modern,candy2016bayesian}. Based on such criterion, we employed the EM algorithm as $C(\cdot)$, which is an efficient parameter estimation algorithm commonly used for GMM \cite{dempster1977maximum,balakrishnan2017statistical}.
	The complexity of \algref{GMM algorithm} is dominated by estimating ${\widehat{\mathcal{P}}}_{\rm I}$, ${\widehat{\mathcal{P}}}_{\rm C}$, and ${\widehat{\mathcal{P}}}_{\rm L}$ (step 3), as well as the LLR computation (steps 5-10).
	The computational complexity of the former (EM algorithm) is $\mathcal{O}(N_{\rm g}{L_{\rm P}})$ in every iteration. For each ${\bf y}^{\rm d}[l]$, steps 5-10 have a complexity order of $\mathcal{O}(N_{\rm g}Q)$. The computational complexities of GMM is summarized in Table \ref{complexity comparison}, where the number of iterations is denoted by $\eta_{\rm g}$. Specifically, we set $N_{\rm g}=5$ for the GMM in PFEN and GMM soft demodulators.
	
	\subsubsection{Computational Complexity of PFEN Soft Demodulator}
	We denote $D_{\rm SA}$ and $N_{\rm H}$ as the last dimension of ${\bf Q}$/${\bf K}$/${\bf V}$ and the number of heads in FEB. The dimensions of the matrices ${\bf E}_1$, ${\bf E}_2$, and ${\bf M}$ are represented by $N_{{\bf E}_1}\times D_{\rm SA}$, $N_{\rm g}\times D_{\rm SA}$, and ${L_{\rm P}}\times D_{{\bf M}}$. Given $N_{{\bf E}_1}\gg N_{\rm g}$, the four full connection layer in FEB have the complexity of $\mathcal{O}(N_{{\bf E}_1}D^2_{\rm SA}+{L_{\rm P}}{D_{{\bf M}}}D_{\rm SA})$. `Multi-Head Attention' module has the complexity order of $\mathcal{O}(N_{{\bf E}_1}D_{\rm SA}{L_{\rm P}})$ \cite{vaswani2017attention}. Thus, PFEN soft demodulator computates ${\widehat{\mathcal{P}}}_{\rm I}$, ${\widehat{\mathcal{P}}}_{\rm C}$, and ${\widehat{\mathcal{P}}}_{\rm L}$ with complexity of $\mathcal{O}({L_{\rm P}}({D_{{\bf M}}}D_{\rm SA}+N_{{\bf E}_1}D_{\rm SA})+N_{{\bf E}_1}D^2_{\rm SA})$. According to Table \ref{table-submodel}, the parameter dimensions are set as $D_{\rm SA}=32$, $N_{\rm H}=4$, $N_{{\bf E}_1}=16$.
	
	The computational complexities of soft demodulators are summarized in Table \ref{complexity comparison}, where the number of iterations in CA is denoted by $\eta_{\rm ca}$.
	The computational complexity of LLR computation (steps \ref{llr start}-\ref{llr end} in \algref{GMM algorithm}) is ignored since it depends on $L_{\rm D}$ and is and is relatively smaller compared to that of estimations, and the demodulation performance is determined by ${\widehat{\mathcal{P}}}_{\rm I}$, ${\widehat{\mathcal{P}}}_{\rm C}$, and ${\widehat{\mathcal{P}}}_{\rm L}$. Note that
	\begin{align}
		{L_{\rm P}}>\eta_{\rm ca}>\eta_{\rm g}\gg D_{\rm SA} > M_{\rm ca}>N_{{\bf E}_1} > N_{\rm g}\geq D_{{\bf M}}.
	\end{align}
	The average number of multiplications is computed with $N=K=8$ and $N_{\rm g}=5$, and the average execution time is evaluated on Intel Xeon W-2150B CPU (3.00GHz). The configuration of QAM and PSK are 1) ASM precoder, 16QAM, $L_{\rm P}=1024$, and ${\rm SNR}=0,5,...,40$; 2) CISB-PA precoder, 16PSK, $L_{\rm P}=128$, and ${\rm SNR}=0,5,...,40$. PFEN has a greater advantage in execution time since it contains no iterative computations.
	
	\subsection{Coded System Performance}\label{coded system performance}

	In this subsection, we consider the coded system that employs the low-density parity check (LDPC) coding scheme \cite{1057683}. 
	In this section, \textbf{Gaus} and \textbf{CA} represent Gaussian and CA soft demodulators, while \textbf{MGaus}, \textbf{GMM}, and \textbf{PFEN} represent the modified Gaussian, GMM, and PFEN demodulators proposed in Sections \ref{MGaus demodulator}, \ref{GMM demodulator}, and \ref{PFEN demodulator}, respectively. 
	Since $L_{\rm P}$ (the number followed by the demodulator in the legends) has little impact on \textbf{Gaus} and \textbf{MGaus}, the simulation focuses on the effect of $L_{\rm P}$ on \textbf{GMM} and \textbf{PFEN}.
	Except for the spectrum efficiency\footnote{The computation method of spectrum efficiency refers to Section \uppercase\expandafter{\romannumeral4} in \cite{9910472}.}, we use the mutual information (MI) between the coded bits ${\bf B}$ and the corresponding LLR, which is approximated using the following formulation, to comprehensively evaluate the performance of the SLP transceivers \cite{Brink2001, el2013exit}. 
	\begin{align}
		\begin{split}	
			&{I({\bf B};{\rm {\bf LLR}})} \approx \\
			&1-\frac{1}{L_{\rm D}\cdot{\rm log}_{2}Q}\sum\limits^{L_{\rm D}}_{l=1}\sum\limits^{{\rm log}_{2}Q}_{i=1}{\rm log}_2\left[1+\exp\left({-{b_i[l]}\cdot{{\widehat{\rm LLR}}_i[l]}}\right)\right],
			\label{E7}
		\end{split}
	\end{align}
	where $b_i[l]$ denotes the $i$-th coded bit of ${\bf y}[l]$.
	
%
	Fig. \ref{recieved distribution} shows the normalized $f_{{ Y}}({{\bf y}}[l]|{\bf{v}}_q)$ with ZF, CIMMSE, and ASM precodings. Different from the circular density plot of the Gaussian distribution, the ASM exhibits an elliptical distribution center in Fig.\ref{recieved distribution} (a3), and there are scattered irregular points away from the decision region in both Fig.\ref{recieved distribution} (a3) and (b3), which can impact the demodulation of the received signals.
	
	\begin{figure}[htp]
		\centering
		\subfigure[16QAM, $N=K=8$.]{
			\includegraphics[width=3in]{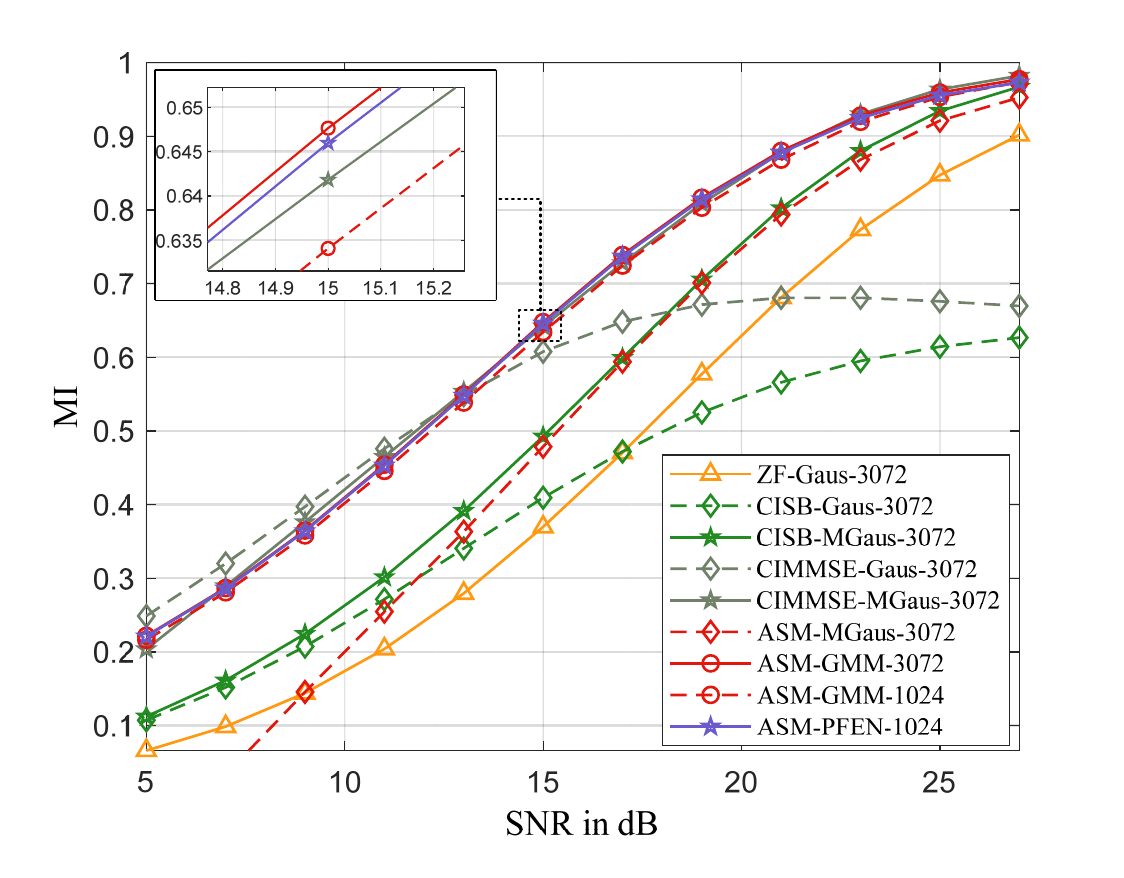}
			\label{MI_8mul8_16qam}
		}
		\subfigure[64QAM, $N=K=12$.]{
			\includegraphics[width=3in]{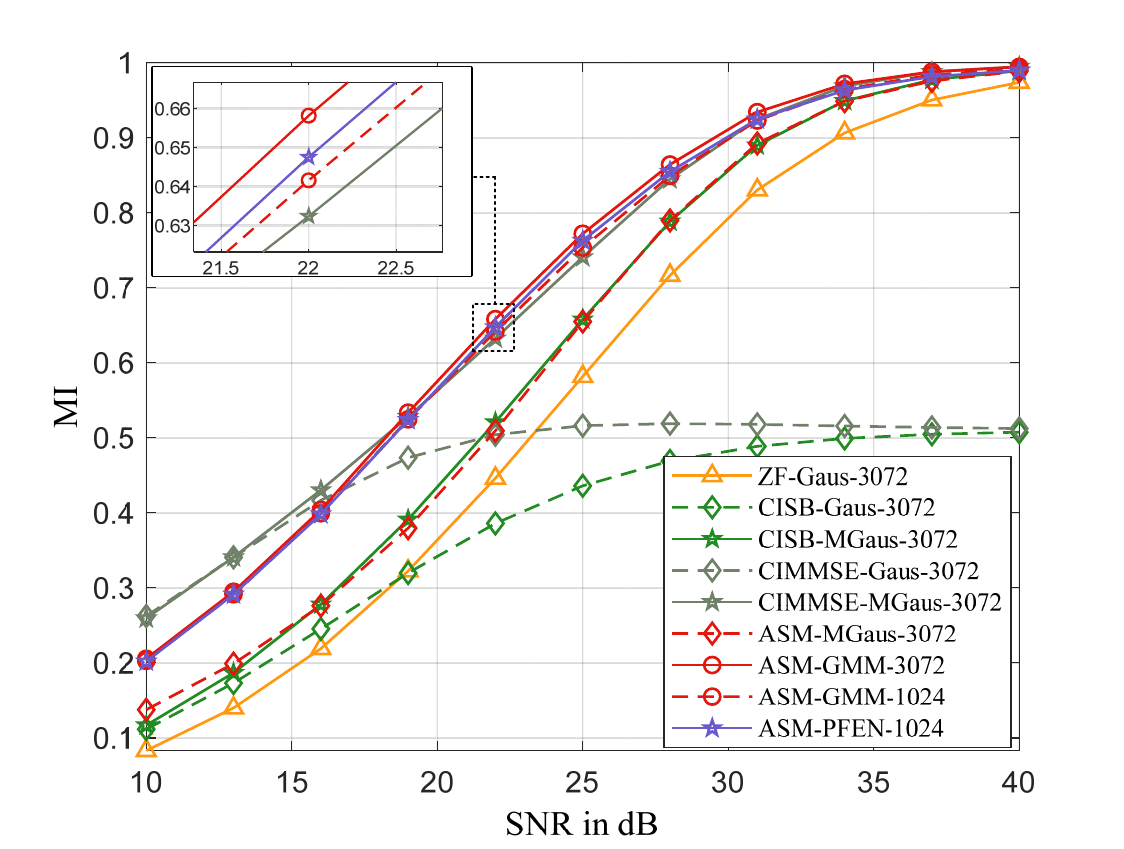}
			\label{MI_12mul12_64qam}
		}
		\DeclareGraphicsExtensions.
		\caption{MI vs SNR, QAM, $L_{\rm D}=2048$.}
		\label{MI_qam}
	\end{figure}
	\begin{figure}[!htb]
		\centering
		\includegraphics[width=3in]{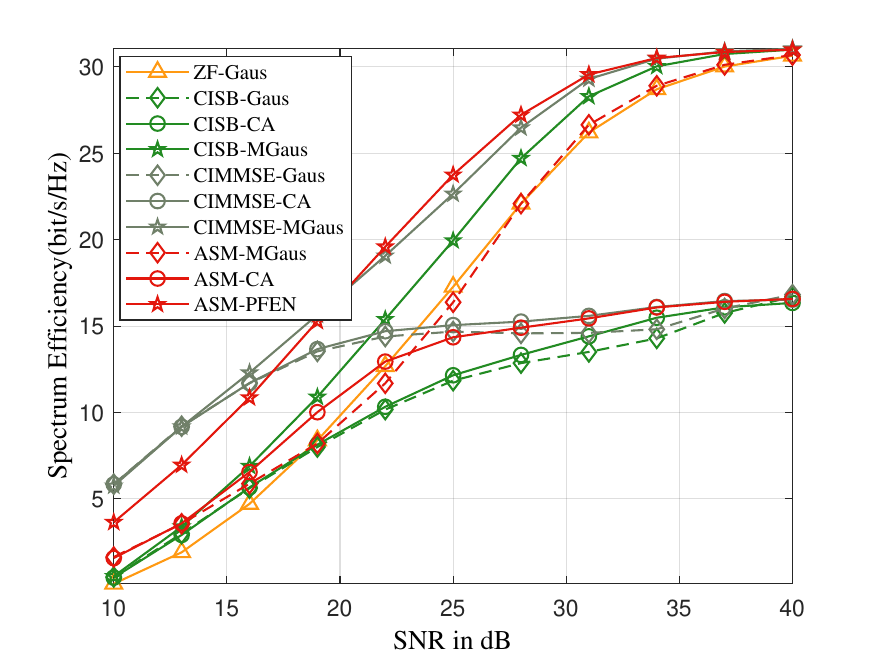}
		\caption{Spectrum Efficiency vs SNR, 64QAM, $N=K=12$, $L_{\rm P}=1024$, $L_{\rm D}=2048$, LDPC.}
		\label{tp}
	\end{figure}
	Fig. \ref{MI_8mul8_16qam} and Fig. \ref{MI_12mul12_64qam} compare MI performance between different transceivers. 
	Compared with CIMMSE-Gaus-3072, which levels off at about 0.65 for 16QAM and further degrades for 64QAM, the MI of CIMMSE-MGaus-3072 shows significant improvement, reaching 1.00 at higher SNR values. The similar trend can also be observed from the performance of CISB with demodulators. The performance of transceiver ASM-PFEN-1024 is almost the same as ASM-GMM-3072, which indicates that \textbf{PFEN} can effectively reduce the pilot overhead.  
	When ${\rm MI}\!=\!0.5$ in scenarios $N \!=\! K \!=\! 8$ with 16QAM and $N \!=\! K \!=\! 12$ with 64QAM, 
	ASM-GMM-1024 transceiver provides SNR gains of about 2.7dB and 3.7dB than ASM-MGaus-3072, and ASM-PFEN-1024 transceiver provides SNR gains of about 3.1dB and 4.0dB. The performance gap between the \textbf{GMM} and \textbf{MGaus} demodulators under the ASM scheme illustrates the significant impact of non-Gaussian signals from ASM on the LLR calculation of \textbf{MGaus}. Due to the similar performance of \textbf{CA} to \textbf{Gaus} and the abundance of curves, we have omitted the MI performance of \textbf{CA} in Fig. \ref{MI_qam} for readability.
	
	\begin{figure}[htb]
		\centering
		\subfigure[CISB, WOR.]{
			\includegraphics[width=3in]{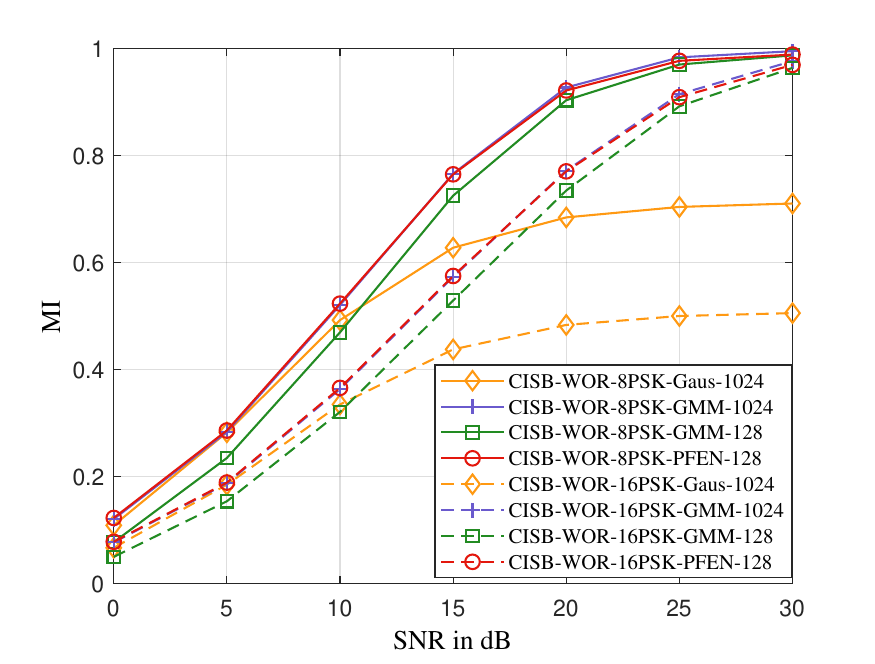}
			\label{CISB-NPA-MI}
		}
		\subfigure[CIMMSE, WOR.]{
			\includegraphics[width=3in]{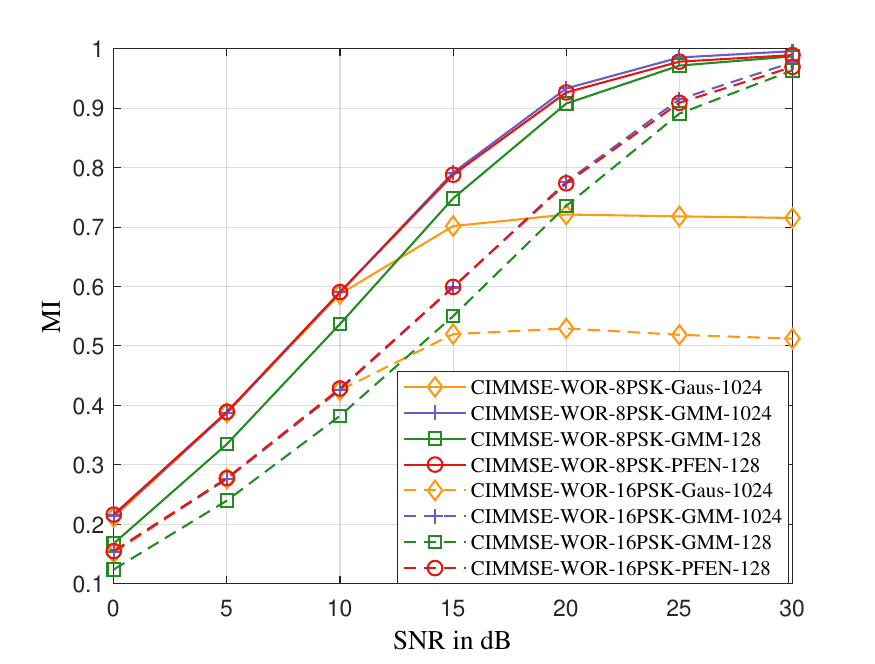}
			\label{CIMMSE-NPA-MI}
		}
		\subfigure[16PSK, $L_{\rm P}=128$.]{
			\includegraphics[width=3in]{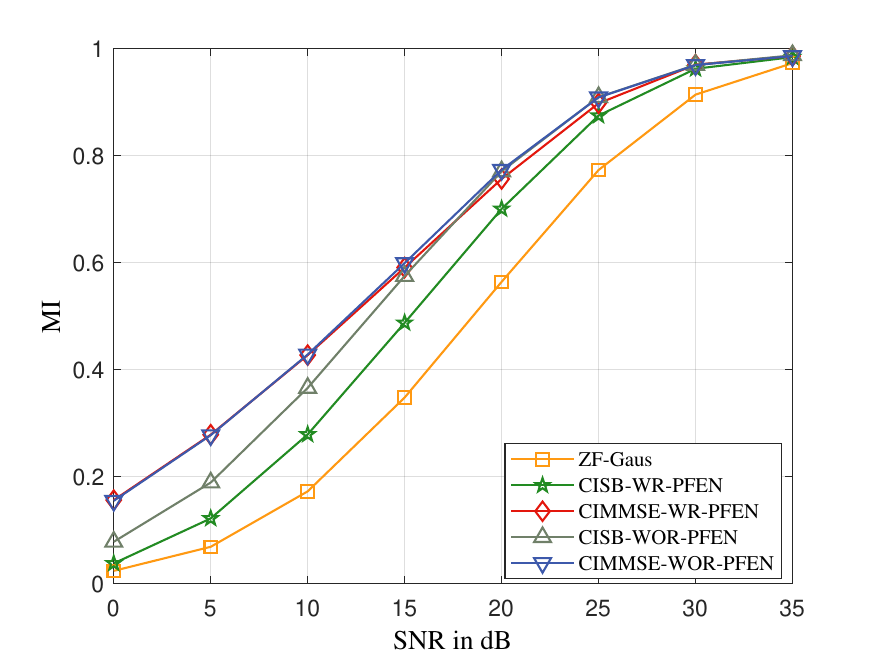}
			\label{16PSK-MI}
		}
		\DeclareGraphicsExtensions.
		\caption{MI vs SNR, PSK, $N=K=8$, $L_{\rm D}=2048$.}
		\label{PSK-MI}
	\end{figure}
	Fig. \ref{tp} shows a comparison of the spectrum efficiency for 64QAM and $N=K=12$. It is evident that the Gaussian soft demodulator severely limits the performance of CISB and CIMMSE, while \textbf{MGaus} provides outstanding throughput for these schemes. Since \textbf{MGaus} does not match the distribution of received signals from ASM, the spectrum efficiency of ASM-PFEN is significantly higher than ASM-MGaus. When the spectrum efficiency is 15 bits/s/Hz, ASM-PFEN provides an SNR gain of about 3.3dB than CISB-MGaus. Moreover, ASM-PFEN outperforms CIMMSE-MGaus slightly in high SNR regimes. 
	It is worth noting that the spectrum efficiency and MI performance of a transceiver, such as CISB-Gaus, may show some differences due to their different sensitivities to LLR distribution. \textbf{CA} cannot provide excellent demodulation performance since its model cannot adapt to the signals with SLP, and it also faces the variance estimation issue as \textbf{Gaus}.
	
	\begin{figure}[tb]
		\centering
		\subfigure[CISB.]{
			\includegraphics[width=3in]{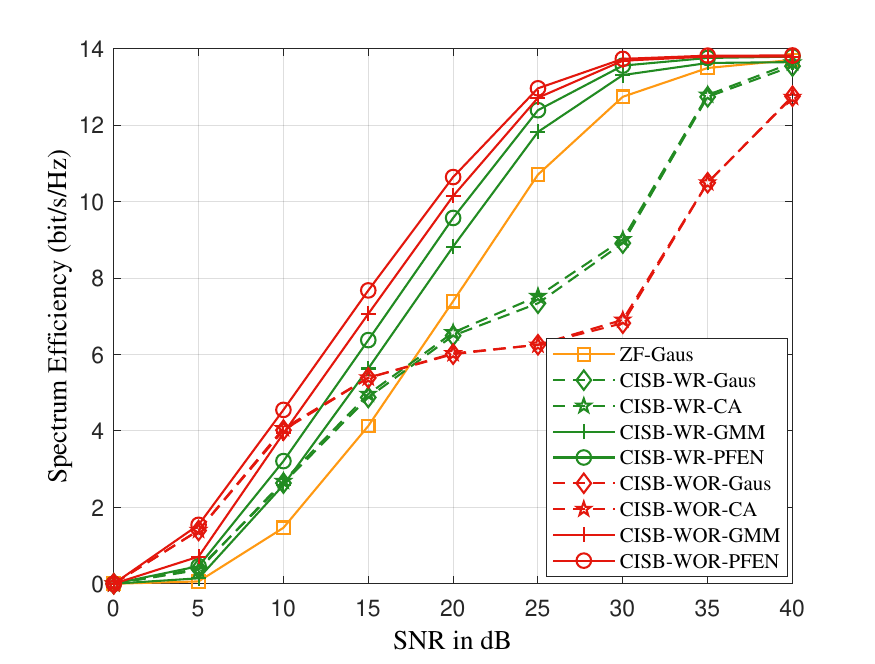}
			\label{CISB Spectrum Efficiency PSK}
		}
		\subfigure[CIMMSE.]{
			\includegraphics[width=3in]{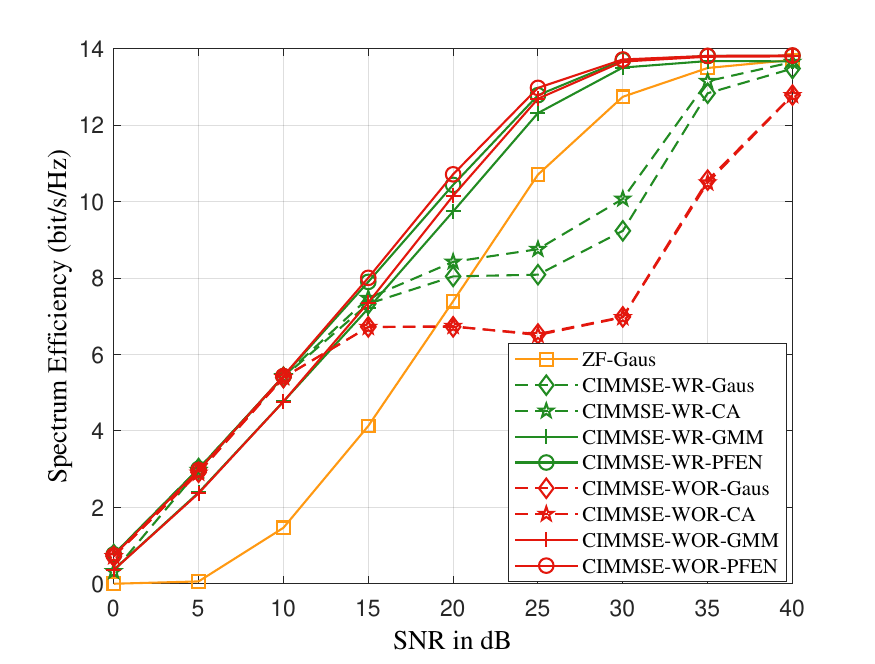}
			\label{CIMMSE Spectrum Efficiency PSK}
		}
		\subfigure[16PSK.]{
			\includegraphics[width=3in]{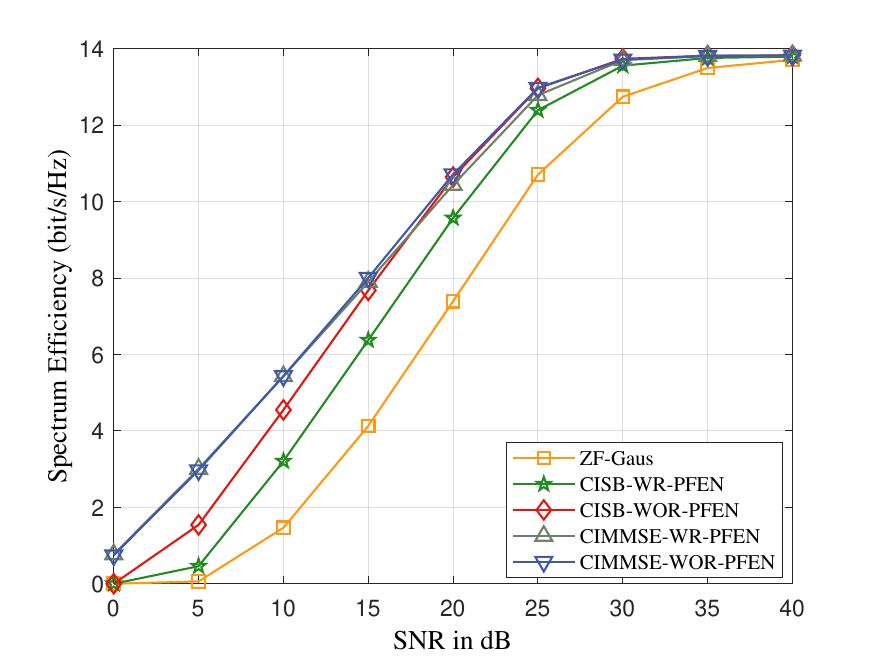}
			\label{Spectrum Efficiency 16PSK}
		}
		\DeclareGraphicsExtensions.
		\caption{Spectrum Efficiency vs SNR, 16PSK, $N=K=8$, $L_{\rm P}=128$, $L_{\rm D}=2048$, LDPC.}
		\label{Spectrum Efficiency PSK}
	\end{figure}
	
	The MI comparison for PSK transmission is depicted in Fig. \ref{PSK-MI}. In Fig. \ref{CISB-NPA-MI}, CISB-NPA-Gaus for 8PSK and 16PSK could only reach MI of about 0.72 and 0.51 when SNR is 30dB, while the MIs of \textbf{GMM} and \textbf{PFEN} grow with the increasing SNR and reach 1.00. When $L_{\rm P}$ decreases from 1024 to 128, \textbf{PFEN} achieves the same performance as \textbf{GMM} with $L_{\rm P}=1024$, while \textbf{GMM} has a significant drop in performance. Since CIMMSE has a similar received signal distribution, the above analysis is also applicable to CIMMSE-NPA in Fig. \ref{CIMMSE-NPA-MI}. It is shown in Fig. \ref{16PSK-MI} that SLP schemes outperform ZF precoding scheme with the help of \textbf{PFEN} soft demodulator, where CIMMSE-NPA-FPEN provides a $72\%$ MI gain than ZF-Gaus when SNR is 15dB.
	
	Fig. \ref{Spectrum Efficiency PSK} illustrates the comparison of spectrum efficiency for PSK transmission. The spectrum efficiency of CISB-NPA-Gaus and CISB-PA-Gaus is much lower than `ZF-Gaus' in the high SNR regime. CISB-NPA-PFEN and CISB-PA-PFEN provide spectrum efficiency gains of about $46\%$ and $75\%$ than ZF-Gaus.
	While CISB-NPA-PFEN has better performance than CISB-PA-PFEN in Fig. \ref{CISB Spectrum Efficiency PSK}, the performance of CIMMSE-NPA-PFEN and CIMMSE-PA-PFEN in Fig. \ref{CIMMSE Spectrum Efficiency PSK} is very similar. In Fig. \ref{Spectrum Efficiency PSK}, all these excellent transceivers are compared, and CIMMSE-NPA-PFEN has the highest spectrum efficiency. 
	
	\section{Conclusion}\label{conclusion}
	This paper investigated the non-Gaussian soft demodulator for SLP in a coded MU-MISO system. 
	We first analyzed the non-Gaussian characteristics of both PSK and QAM signals with existing SLP schemes and categorized the non-Gaussian signals into two distinct types.
	To achieve precise LLR estimation from the two categories of non-Gaussian received signals, we put forward the modified Gaussian and the GMM soft demodulators. Additionally, we proposed the PFEN demodulator based on the transformer mechanism in deep learning, which effectively reduces computational complexity and pilot overhead. Simulation results demonstrated that the proposed soft demodulators significantly enhance the throughput of existing SLPs for both PSK and QAM transmission in coded systems.

	
	%

	\appendices
		
	\section{Proof of Proposition \ref{rotation ppn}}\label{ppn 1 proof}
	Without loss of generality, we present the proof for the case of PSK transmission with CISB. The problem of CISB is given by \cite{8299553,8466792}
	\begin{align}
		\begin{split}
			&\max\limits_{{\bf x},\gamma} \
			\gamma	
			\\
			&s.t.~  {\bf h}^T_k{\bf x}\in \gamma\cdot\mathcal{D}_k,\;\forall k\in\mathcal{K},\\
			& \qquad \|{\bf x}\|^2_2=P_{\rm T}.
		\end{split}
		\label{CI-ZF formulation}
	\end{align}
	Given $\{{\bf h}^T_k\}^K_{k=1}$ and $P_{\rm T}$, the optimal ${\bf x}^{\star}$ and $\gamma^{\star}$ are determined by ${\bf s}$, thus we use ${\bf x}^{\star}({\bf s})$ and $\gamma^{\star}({\bf s})$ to denote them. As in Section \ref{S4}, we consider the received signals of $k$-th UE. According to \eqref{E1}, and \eqref{PSK demodulation}, ${\bar{y}}$ is given by
	\begin{align}
		{\bar{y}} = 
		\begin{cases}
			\frac{{\bf h}_k^T{\bf x}^{\star}({\bf s})}{\gamma^{\star}({\bf s})}+\frac{n_k}{{\bar \gamma}}, \text{WR}\\
			{\bf h}_k^T{\bf x}^{\star}({\bf s})+n_k, \text{WOR}
		\end{cases}.
	\end{align}
	We focus on the proof for the case with power allocation, which can be extended to that without power allocation.
	Since symbols in ${\mathcal{V}}$ are transmitted with equal probability, ${\bf s}$ follows a discrete uniform distribution with the sample space $\mathcal{V}^K$. 
	Furthermore, ${\bf s}_q,\forall q \in \mathcal{Q}$ is defined as the discrete uniformly distributed variable with the following sample space:
	\begin{align}
	{\mathcal{S}}_q = \{{\bf s}|{\bf s}\in\mathcal{V}^K,[{\bf s}]_k=v_q\},\ \forall q \in \mathcal{Q}. 
	\label{sub S space}
	\end{align}	
	We respectively denote ${\bf s}^{{\rm sp}}_m$ and ${\bf s}^{{\rm sp}}_n$ as the observations of ${\bf s}_m$ and ${\bf s}_n$. For PSK transmission, it can be proved that the following equation is a bijection between ${\mathcal{S}}_m$ and ${\mathcal{S}}_n$
	\begin{align}
		{\bf s}^{{\rm sp}}_m = {\bf s}^{{\rm sp}}_n\cdot\exp[j(m-n)\frac{2\pi}{Q}],\ \forall m,n\in {\mathcal{ Q}}.
		\label{rotation group}
	\end{align}
	The above bijection also holds for $m$ and $n$ satisfying $\frac{m-n}{4}\in{\mathbb{Z}}$ in 16QAM transmission.
	
	Based on the definition of $\mathcal{D}_k$ and the convexity of problem \eqref{CI-ZF formulation}, it can be verified that 
	\begin{align}
	{\bf x}^{\star}({\bf s}^{{\rm sp}}_m) &= {\bf x}^{\star}({\bf s}^{{\rm sp}}_n)\cdot\exp[j(m-n)\frac{2\pi}{Q}],\\
	\gamma^{\star}({\bf s}^{{\rm sp}}_m)&=\gamma^{\star}({\bf s}^{{\rm sp}}_n),
	\end{align}
	where ${\bf s}^{{\rm sp}}_m$ and ${\bf s}^{{\rm sp}}_n$ are observations satisfying \eqref{rotation group}. Furthermore, we have
	\begin{align}
		\frac{{\bf h}_k^T{\bf x}^{\star}({\bf s}^{{\rm sp}}_m)}{\gamma^{\star}({\bf s}^{{\rm sp}}_m)} = \frac{{\bf h}_k^T{\bf x}^{\star}({\bf s}^{{\rm sp}}_n)}{\gamma^{\star}({\bf s}^{{\rm sp}}_n)}\cdot\exp[j(m-n)\frac{2\pi}{Q}].\label{rotation equation2}
	\end{align}
	Although the problem of ASM is non-convex, it can be proven the variables after each iteration of Algorithm 1 in \cite{Wang2023} satisfy \eqref{rotation equation2}.
	
	According to \eqref{sub S space}, \eqref{rotation group}, and \eqref{rotation equation2}, $\frac{{\bf h}_k^T{\bf x}^{\star}({\bf s}_m)}{\gamma^{\star}({\bf s}_m)}$ are identically distributed with $\frac{{\bf h}_k^T{\bf x}^{\star}({\bf s}_n)}{\gamma^{\star}({\bf s}_n)}\cdot\exp[j(m-n)\frac{2\pi}{Q}]$. Since $n_k$ follow $\mathcal{C}\mathcal{N}(0, \sigma^2)$ that is circular symmetric, ${\bar{y}}_m$ and ${\bar{y}}_n\cdot\exp[j(m-n)\frac{2\pi}{Q}]$ are identically distributed, where ${\bar{y}}_m$ and ${\bar{y}}_n$ represent ${\bar{y}}$ with transmit symbol vector ${\bf s}_m$ and ${\bf s}_n$, respectively. As  ${f}_{{ Y}}\left({\bf y}|{\bf{v}}_q\right)$ is the PDF of $\begin{bmatrix}
			\real\left({\bar { y}}_q\right)\\
			\imaginary\left({\bar { y}}_q\right)
		\end{bmatrix}$, we have
	\begin{align}
		{f}_{{ Y}}\left({\bf y}|{\bf{v}}_m\right) = {f}_{{ Y}}\left(R_{m-n}({\bf y})|{\bf{v}}_n\right).
	\end{align}	
	This concludes the proof.
	
	\bibliographystyle{IEEEtran}
	\bibliography{Refabrv,IEEEfull}

\end{document}